\pdfoutput=1
\documentclass[aps,prd,floatfix,twocolumn,reprint,superscriptaddress]{revtex4-1}
\usepackage{amsfonts}
\usepackage{mathrsfs}
\usepackage{amsmath}
\usepackage{color}
\usepackage{graphicx}
\usepackage{bm}
\usepackage{amssymb}
\usepackage{float}
\usepackage{xspace}
\usepackage{epstopdf}
\usepackage{dcolumn}
\usepackage{longtable}
\usepackage[colorlinks=true, letterpaper=true, pdfstartview=FitV, linkcolor=blue, citecolor=blue, urlcolor=blue]{hyperref}

\begin{document}
\title{Nodal line semimetal states in positive electrode material of lead-acid battery: Lead dioxide family and its derivatives}

\author{Run-Wu Zhang}
\affiliation{Beijing Key Laboratory of Nanophotonics and Ultrafine Optoelectronic Systems, School of Physics, Beijing Institute of Technology, Beijing 100081, China}
\affiliation{China Academy of Engineering Physics, Mianyang, Sichuan, 621900, China}

\author{Cheng-Cheng Liu}
\email{ccliu@bit.edu.cn}
\affiliation{Beijing Key Laboratory of Nanophotonics and Ultrafine Optoelectronic Systems, School of Physics, Beijing Institute of Technology, Beijing 100081, China}

\author{Da-Shuai Ma}
\affiliation{Beijing Key Laboratory of Nanophotonics and Ultrafine Optoelectronic Systems, School of Physics, Beijing Institute of Technology, Beijing 100081, China}

\author{Maoyuan Wang}
\affiliation{Beijing Key Laboratory of Nanophotonics and Ultrafine Optoelectronic Systems, School of Physics, Beijing Institute of Technology, Beijing 100081, China}

\author{Yugui Yao}
\email{ygyao@bit.edu.cn}
\affiliation{Beijing Key Laboratory of Nanophotonics and Ultrafine Optoelectronic Systems, School of Physics, Beijing Institute of Technology, Beijing 100081, China}

\begin{abstract}
Based on first-principles calculations and symmetry analysis, we report that the three-dimensional (3D) nodal line (NL) semimetal phases can be realized in the lead dioxide family (\emph{$\alpha$}-PbO$_2$ and \emph{$\beta$}-PbO$_2$) and its derivatives. The \emph{$\beta$}-PbO$_2$ features two orthogonal nodal rings around the Fermi level, protected by the mirror reflection symmetry. The effective model is developed and the related parameters are given by fitting with the HSE06 band structures. The NLs mainly come from the $p$ orbitals of the light element O and are rather robust against such tiny spin-orbit coupling. The NL phase of the \emph{$\alpha$}-PbO$_2$ can be effectively tailored by strain, making a topological phase transition between a semiconductor phase and a NL phase. In addition, the exploration of \emph{$\beta$}-PbO$_2$ derivatives (\emph{i.e.} \emph{$\beta$}-PbS$_2$ and \emph{$\beta$}-PbSe$_2$) and confirmation of their topological semimetallicity greatly enrich the NL semimetal family. These findings pave a route for designing topological NL semimetals and spintronic devices based on realistic PbO$_2$ family.

\end{abstract}
\maketitle
\section{Introduction}
Nodal semimetals ~\cite{1,2,3,4,5,6} characterized by the patterns of band-crossing nodes around the Fermi level have attracted
broad interest recently in condensed matter physics and material science. Based on different patterns of the band-crossing nodes (either zero-dimensional discrete points or one-dimensional continuous lines) in momentum space~\cite{7,8}, nodal semimetals fall into different categories, for instance, Dirac (Weyl) semimetals~\cite{9,10,11,12,13,14,15,16,17,18,19,20,21,22,23,24} and crossing nodal line semimetals~\cite{25,26,27,28,29,30,31,32,33,34,35,36,37,38,39,40,41,42,43,44,45,46,47,48,49,50,51,52,53}. In contrast to Dirac (Weyl) semimetals, the latter with nontrivial line connectivity includes nodal line (NL) semimetals~\cite{25,26,27,28,29,30,31,32,33,34,35,36,37,38,39,40,41}, nodal chain semimetals~\cite{42,43}, nodal link semimetals~\cite{44,45,46}, and other nodal connected semimetals~\cite{47,48,49,50}. These NL semimetals have attracted plenty of interest due to the remarkable physical properties, such as the ultrahigh mobility~\cite{51}, the flat drumhead-like surface states~\cite{25,26,27,28,29,30,31,32,33,34,35,36,37,38,39,40,41,42,43,44,45,46,47,48,49,50}, the long-range Coulomb interaction~\cite{52}, the flat Landau level spectrum ~\cite{53}, the special collective modes~\cite{54}, \emph{etc}.

However, due to the lack of an ideal candidate and its experimental realization, these remarkable properties of NL semimetals are less explored. To date, although some theoretical NL semimetals have been proposed, these materials are often suboptimal and suffer from various drawbacks: such as the nodal line located far away from the Fermi level and entangled with other irrelevant trivial bands, as well as the complex materials hard to implement experimentally. Therefore, searching for realistic materials with ideal NL characters will provide an important platform to investigate such fascinating physical properties in NL semimetals.

In this paper, overcoming all these drawbacks, we theoretically propose that the ideal NL semimetal states can be realized in the PbO$_2$ family (\emph{$\alpha$}-PbO$_2$ and \emph{$\beta$}-PbO$_2$), which are widely uesed as classical positive electrode materials in the lead-acid battery~\cite{55,56,57,58,59,60,61,62,63,64}. These materials have already been synthesized in experiment more than a century and attracted tremendous interest from academy to industry. Unlike other proposed three perpendicular NLs such as Cu$_3$PdN~\cite{65}, LaN~\cite{66}, and CaTe~\cite{67}, we show that the \emph{$\beta$}-PbO$_2$ exhibits two orthogonal nodal rings with a negligible spin-orbit coupling (SOC). Such two nodal rings in \emph{$\beta$}-PbO$_2$ are protected by two perpendicular mirror planes (\emph{i}.\emph{e}., (110) plane and ($1\bar{1}0$) plane). A two-band model is developed for this system. Besides, the hallmark drumhead-like surface states are presented explicitly. Under the hydrostatic strain, topological phase transitions between a semiconductor phase and a topological nodal line (TNL) phase can be triggered in both \emph{$\beta$}-PbO$_2$ and \emph{$\alpha$}-PbO$_2$. With such weak SOC, the NL state in \emph{$\beta$}-PbO$_2$ is driven into a topological Dirac semimetal state protected by fourfold screw rotation ($\tilde{C}_{4z}$) symmetry. Moreover, the prediction of \emph{$\beta$}-PbX$_2$ derivatives (\emph{i}.\emph{e}., \emph{$\beta$}-PbS$_2$ and \emph{$\beta$}-PbSe$_2$) and the conformation of their semimetallicity have greatly enriched the NL semimetal family.


To obtain accurate structures of PbO$_2$ family, a recently developed van der Waals (vdW) density functional (\emph{i}.\emph{e}., SCAN+rVV10) is employed~\cite{68}. The SCAN+rVV10 method can reach the accuracy comparable to that of higher-level methods such as random-phase approximation (RPA), while the computational cost can be reduced significantly. The first-principles calculations are performed by using the Vienna \emph{ab} \emph{initio} simulation package (VASP) based on the generalized gradient approximation (GGA) in the Perdew-Burke-Ernzerhof (PBE) ~\cite{69} functional and the projector augmented-wave (PAW) pseudopotential ~\cite{70}. The cutoff energy is set as 500 eV. During the structural optimization of PbO$_2$ family, all atomic positions and lattice parameters are fully optimized. The energy and force convergence criteria are set to be 10$^{-5}$ eV and 0.001 eV/$\mathring{\mathrm{A}}$, respectively. The screened exchange hybrid density functional HSE06 ~\cite{71,72} is adopted to further correct the electronic structure. Here we construct the maximally localized Wannier functions (MLWFs) by employing the Wannier90 code ~\cite{73}. The surface state calculations are implemented in WannierTools package ~\cite{74} based on the Green' s functions method.

\section{Crystal structure of \emph{$\alpha$}-PbO$_2$ and \emph{$\beta$}-PbO$_2$}

\begin{figure}
\includegraphics[width=1.0\columnwidth]{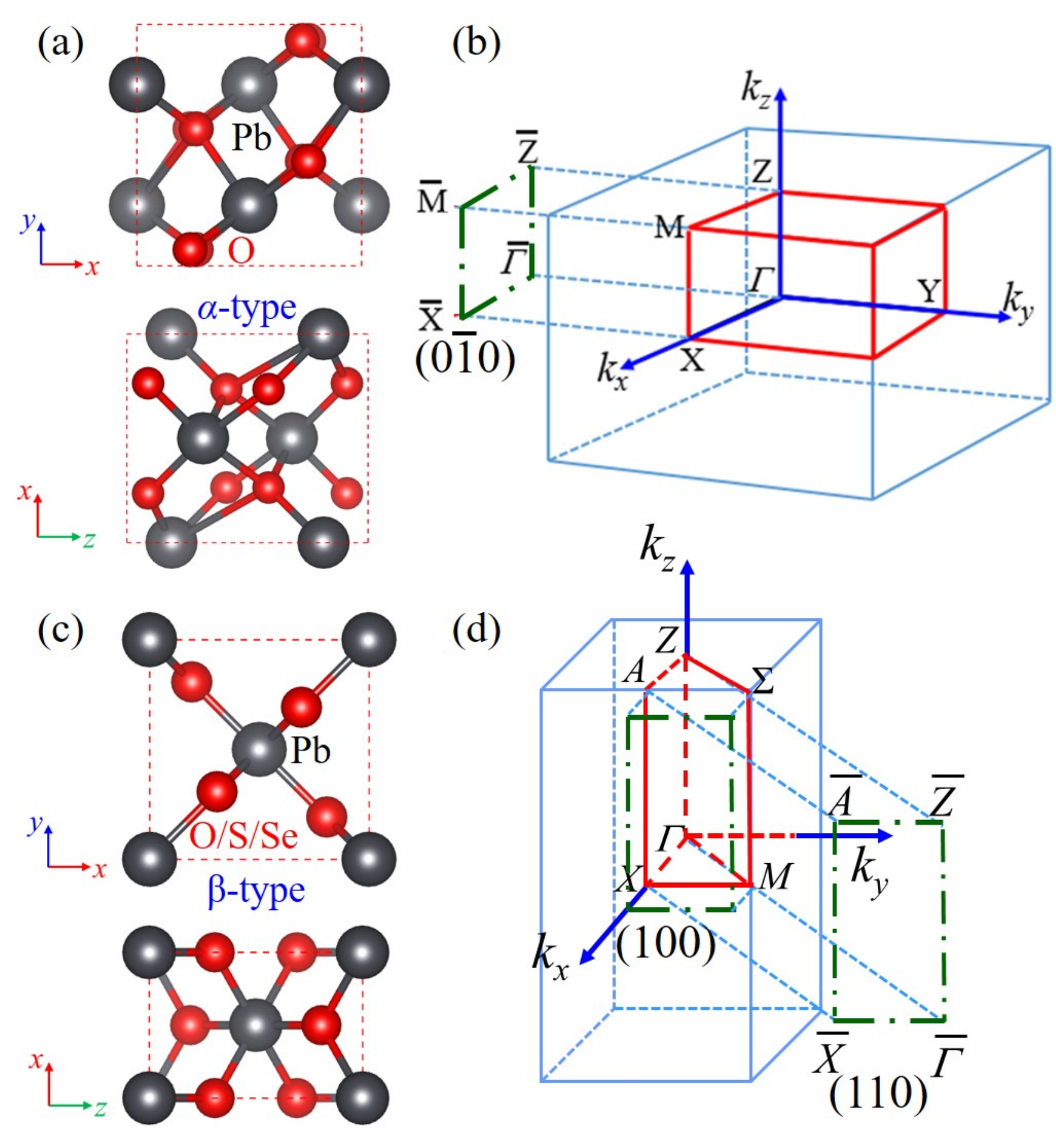}
\caption
{(Color online) (a, c) Top and side views of \emph{$\alpha$}-PbO$_2$ and \emph{$\beta$}-PbX$_2$ (X=O, S, Se) crystals. (b) The 3D Brillouin zone of \emph{$\alpha$}-PbO$_2$ and its projection on the surface Brillouin zone of the ($0\bar{1}0$) surface. (d) The 3D Brillouin zone of \emph{$\beta$}-PbX$_2$ and its projection on the surface Brillouin zone of the (100) and (110) surfaces.}
\label{fig1}
\end{figure}

Lead dioxide (PbO$_2$), as a positive electrode material of lead-acid battery, usually crystallizes in ample phases under different experimental circumstances~\cite{56,57} such as \emph{$\alpha$}-PbO$_2$ and \emph{$\beta$}-PbO$_2$. The \emph{$\alpha$}-PbO$_2$ possesses an orthorhombic structure with the nonsymmorphic space-group \emph{Pbcn} ($D_{2h}^{14}$, No. 60), and the \emph{$\beta$}-PbO$_2$ shares a tetragonal rutile structure with the nonsymmorphic space-group \emph{P42/mnm} ($D_{4h}^{14}$, No. 136), as illustrated in Fig.~\ref{fig1}(a) and Fig.~\ref{fig1}(c). To optimize the structural configuration of \emph{$\alpha$}-PbO$_2$ and \emph{$\beta$}-PbO$_2$ more precisely, the more advanced SCAN+rVV10 method is applied besides the PBE one. The calculated detailed structural parameters are summarized in Table \ref{Table1}. Remarkably, we find that the calculated structural parameters of \emph{$\alpha$}-PbO$_2$ and \emph{$\beta$}-PbO$_2$ by the SCAN+rVV10 method are in excellent agreement with the experimental results ~\cite{61}. Hence, we take the lattice constants of \emph{$\alpha$}-PbO$_2$ and \emph{$\beta$}-PbO$_2$ by the SCAN+rVV10 method to further study in our work hereafter.

\begin{table}
  \caption{The crystal lattice constants (in $\mathring{\mathrm{A}}$) of \emph{$\alpha$}-PbO$_2$ and \emph{$\beta$}-PbO$_2$ are calculated based on the PBE and SCAN+rVV10 methods, respectively. The structural parameters of PbO$_2$ by the SCAN+rVV10 are very close to the experimental results.}
  \label{Table1}
  \begin{tabular}{c|c|c|c}
    \hline
    \hline
    Phase	& Experiment~\cite{61} & PBE   & SCAN+rVV10  \\
            & $a\setminus b\setminus c$ & $a\setminus b\setminus c$ & $a\setminus b\setminus c$\\
    \hline
    \emph{$\alpha$}-PbO$_2$	& $5.00\setminus 5.44\setminus 5.93$ & $5.12\setminus 5.55\setminus 6.04$ & $5.09\setminus 5.47\setminus 5.89$\\
    \hline	
    \emph{$\beta$}-PbO$_2$	& $4.95\setminus 4.95\setminus 3.38$ & $5.09\setminus 5.09\setminus 3.45$ & $4.99\setminus 4.99\setminus 3.42$\\
    \hline
    \hline
  \end{tabular}
\end{table}

\section{Topological nodal line semimetal states in \emph{$\beta$}-PbO$_2$}

\begin{figure}
\includegraphics[width=1.0\columnwidth]{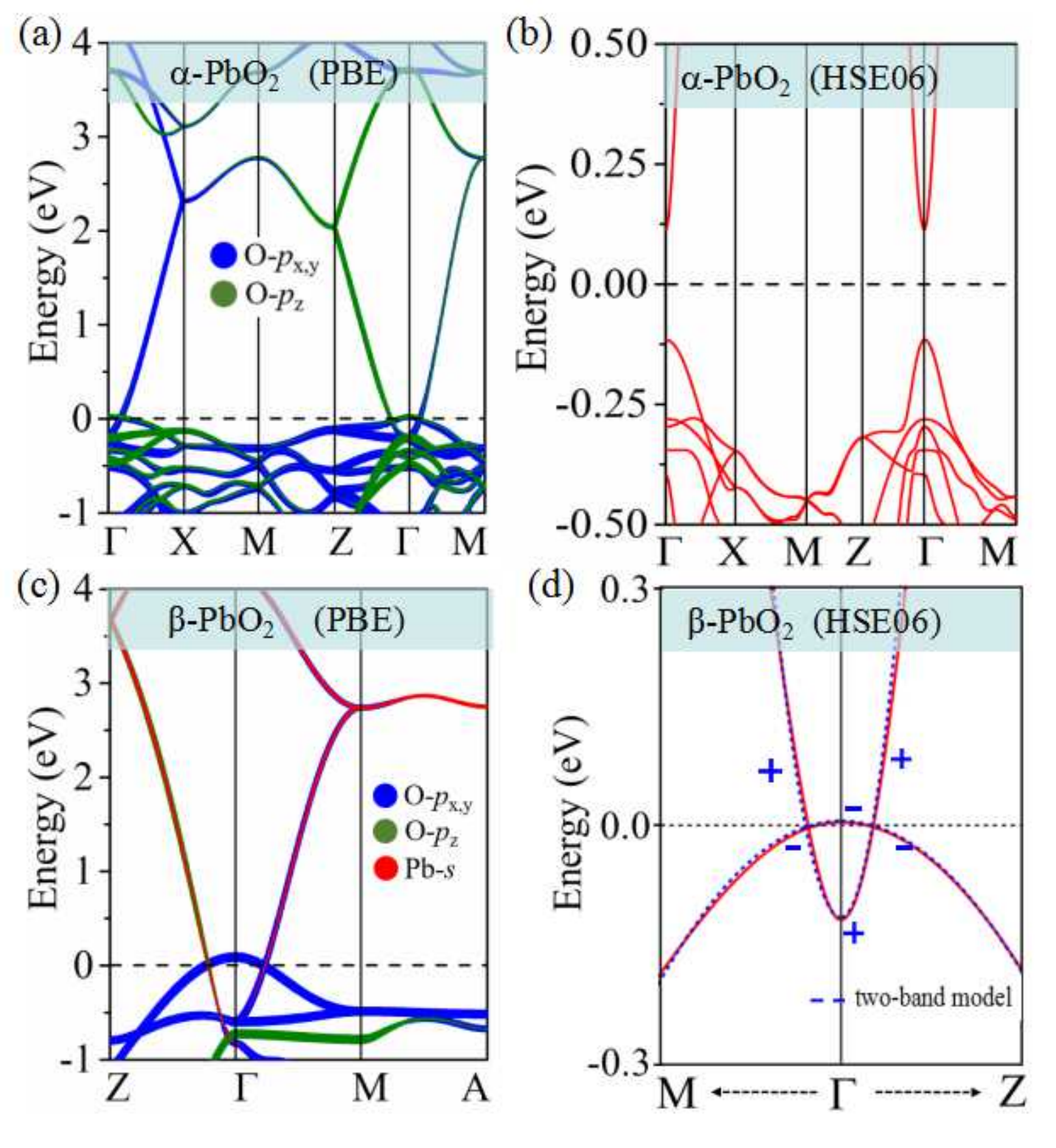}
\caption
{(Color online) The band structures of \emph{$\alpha$}-PbO$_2$ and \emph{$\beta$}-PbO$_2$ without SOC. (a, c) Band structures of \emph{$\alpha$}-PbO$_2$ and \emph{$\beta$}-PbO$_2$ with orbital analysis based on the PBE method. (b, d) Band structures of \emph{$\alpha$}-PbO$_2$ and \emph{$\beta$}-PbO$_2$ are calculated via the HSE06 method. The eigenvalue of mirror reflection symmetry for each band is labeled + and - in (d). The band structures of HSE06 (red solid curves) agrees well with our two-band model (blue dashed curves) calculations.}
\label{fig2}
\end{figure}

We first study the bulk electronic properties of \emph{$\alpha$}-PbO$_2$ and \emph{$\beta$}-PbO$_2$, whose Brillouin zones (BZs) are displayed in Fig.~\ref{fig1}(b) and Fig.~\ref{fig1}(d), respectively. The orbital-projection analysis shows that the obvious contribution to the conduction band minimum (CBM) and valence band maximum (VBM) near the $\Gamma$ point comes from the O-\emph{p}$_x,_y,_z$ orbitals in \emph{$\alpha$}-PbO$_2$ [Fig.~\ref{fig2}(a)], while the CBM and VBM arise mainly from the O-\emph{p}$_x,_y,_z$ orbitals and Pb-\emph{s} orbital in \emph{$\beta$}-PbO$_2$ [Fig.~\ref{fig2}(c)]. Thus such weak SOC effect can be negligible. At the PBE level, several linear band-crossing nodes appear along the high symmetry lines in both phases of PbO$_2$. To further check their linear band-crossing nodes, the more sophisticated HSE06 method is also employed to calculate the band structures, as shown in Fig.~\ref{fig2}(b) and Fig.~\ref{fig2}(d). As can be seen clearly, there are linear band-crossing nodes along the $M \to \Gamma$ and $\Gamma \to Z$ paths in \emph{$\beta$}-PbO$_2$, while the band structure of \emph{$\alpha$}-PbO$_2$ behaves the total characteristic of semiconductors with a direct gap of 0.23 eV under equilibrium state.

\begin{figure}
\includegraphics[width=1.0\columnwidth]{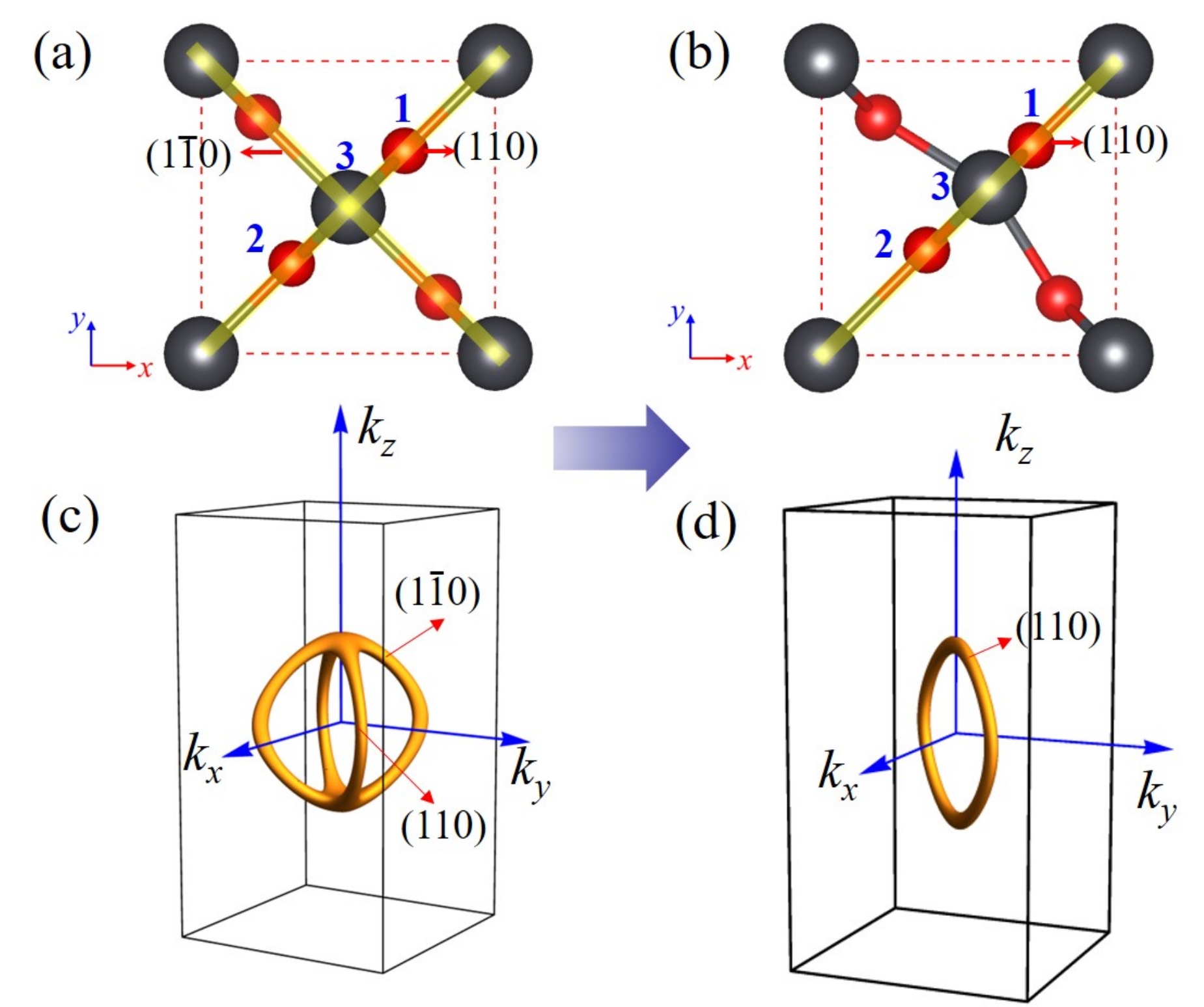}
\caption
{(Color online) (a) Pristine structure of \emph{$\beta$}-PbO$_2$. Two mirror reflection planes $\hat{M}_{xy}$ (110) and $\hat{\tilde{M}}_{xy}$ ($1\bar{1}0$) are marked with yellow zones. (b) Moving O$_1$, O$_2$, and Pb$_3$ atoms aim to break $\mathcal{P}$ and $\hat{\tilde{M}}_{xy}$ ($1\bar{1}0$) mirror symmetries. (c) Two NLs lie on the mutually perpendicular $\hat{M}_{xy}$ (110) and $\hat{\tilde{M}}_{xy}$ ($1\bar{1}0$) mirror planes. (d) One NL lies on the $\hat{M}_{xy}$ (110) mirror plane after breaking $\mathcal{P}$ and ($1\bar{1}0$) mirror symmetries.}
\label{fig3}
\end{figure}

Regarding the space group \emph{P42/mnm} ($D_{4h}^{14}$, No. 136) in \emph{$\beta$}-PbO$_2$, it is important to note the presence of two perpendicular mirror reflection symmetries: $\hat{M}_{xy}$ (110) and $\hat{\tilde{M}}_{xy}$ ($1\bar{1}0$), as plotted in Fig.~\ref{fig3}(a). Since the $M \to \Gamma$ and $\Gamma \to Z$ paths, along which the band-crossing nodes emerge, are on the $\hat{M}_{xy}$ (110) mirror invariant plane, the mirror eigenvalues will be a good quantum number to label the bands on the mirror invariant planes. As shown in Fig.~\ref{fig2}(d), the CBM and the VBM have opposite signs of the eigenvalues with respect to the $\hat{M}_{xy}$ (110) plane, and the band-crossing nodes between these two bands do not just appear at two isolated nodes on the $ M \to \Gamma \to Z$ lines but actually constitute a continuous nodal ring around the $ \Gamma $ point on the $\hat{M}_{xy}$ (110) plane. The similar analysis also applies to the $\hat{\tilde{M}}_{xy}$ ($1\bar{1}0$) mirror plane. Therefore, two nodal rings in \emph{$\beta$}-PbO$_2$ are exactly lying on the mutually perpendicular two mirror invariant planes ($\hat{M}_{xy}$ (110) and $\hat{\tilde{M}}_{xy}$ ($1\bar{1}0$)), as displayed in Fig.~\ref{fig3}(c). This kind of representative NL structure in \emph{$\beta$}-PbO$_2$ is different from the recent predictions of three perpendicular nodal rings such as Cu$_3$PdN~\cite{65}, LaN~\cite{66}, and CaTe~\cite{67}. In addition, the energy range of linear dispersion in \emph{$\beta$}-PbO$_2$ is almost 3 eV (Fig.~\ref{fig2}(c)), much larger than other proposed NL materials. This makes the \emph{$\beta$}-PbO$_2$ could become a promising candidate for investigating the novel physics associated with TNL semimetals.

In general, there are two typical protection mechanisms for the TNL in absence of SOC system, \emph{i}.\emph{e}., mirror symmetry as well as the combination of the spatial inversion ($\mathcal{P}$) and time-reversal ($\mathcal{T}$) symmetries. The configuration of the NLs manifests that \emph{$\beta$}-PbO$_2$ is protected by the mirror symmetry, which is verified by the direct calculation of the eigenvalue of the mirror. This is quite different from the case where the NLs are protected by the $\mathcal{P}\mathcal{T}$ symmetries. Such NLs usually take the special form of the so-called snakelike loop (not necessary to lie on the mirror invariant planes), \emph{e}.\emph{g}., in alkaline-earth compound AX$_2$ (A = Ca, Sr, Ba; X = Si, Ge, Sn)~\cite{75} and triclinic CaAs$_3$~\cite{76}. To further validate our viewpoint, we introduce artificial perturbations to break the inversion symmetry (no $\mathcal{P}\mathcal{T}$ symmetries) while keeping the one mirror symmetry ($\hat{M}_{xy}$ (110) mirror) unchanged. This method involves moving O$_1$, O$_2$, and Pb$_3$ atoms while retaining other atoms, as shown in Fig.~\ref{fig3}(a) and Fig.~\ref{fig3}(b). Our calculated result shows that the NL lying on the $\hat{M}_{xy}$ (110) mirror plane is intact, as shown in Fig.~\ref{fig3}(d). These confirm that the two mutually perpendicular NLs in \emph{$\beta$}-PbO$_2$ are protected by the mirror symmetries rather than the $\mathcal{P}\mathcal{T}$ symmetries.

\begin{figure}
\includegraphics[width=\columnwidth]{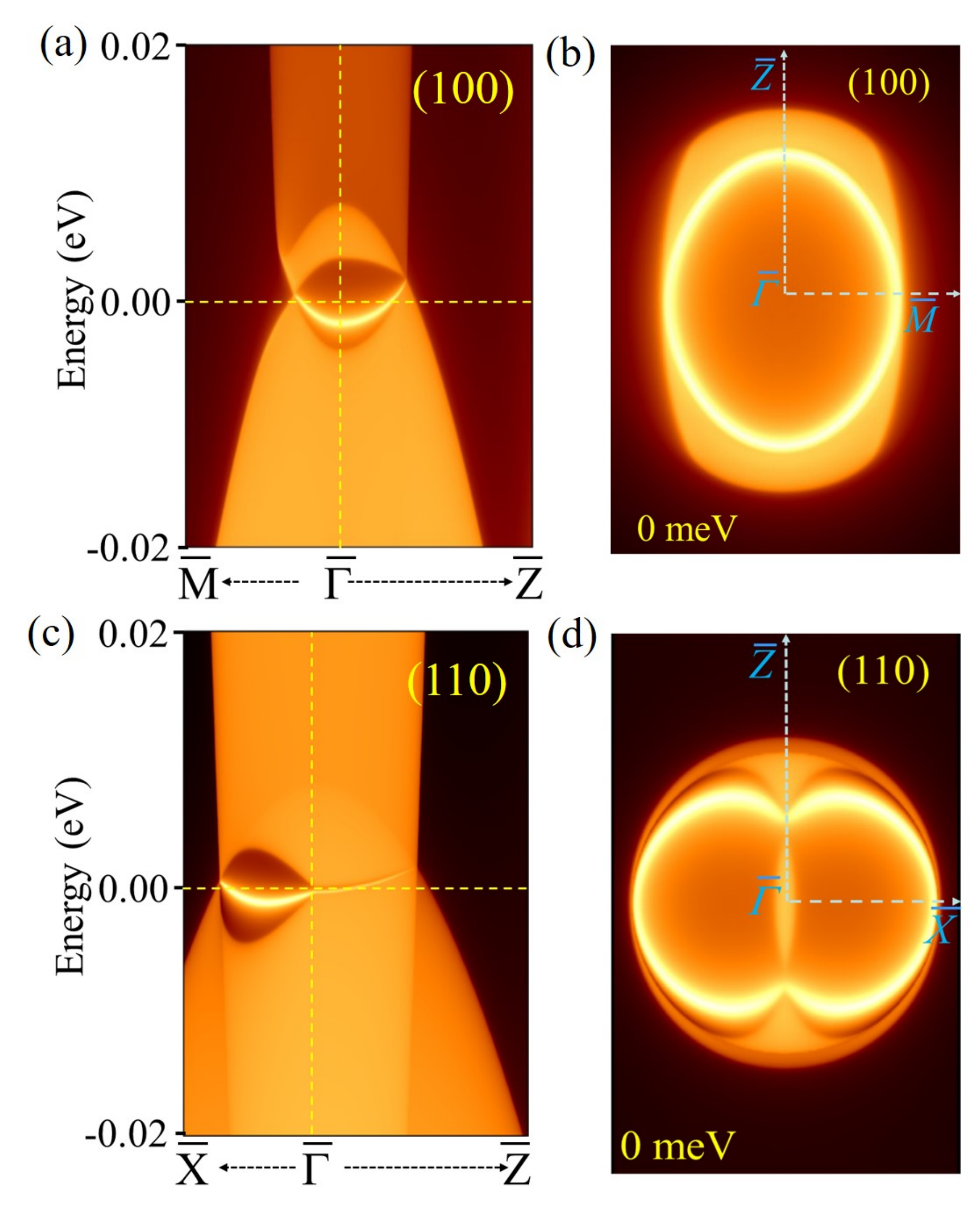}
\caption
{(Color online) (a, c) The projected surface states for the (100) and (110) surfaces. (b, d) The isoenergy surface around the $\Gamma$ point at the Fermi level for the (100) and (110) surfaces.}
\label{fig4}
\end{figure}

In the following, we will address the topological properties of the \emph{$\beta$}-PbO$_2$. The Berry phase along a closed path \emph{C} threading a nodal ring in 3D NL semimetals with $\mathcal{P}\mathcal{T}$ symmetries can be taken as a topological invariant to describe their topological properties~\cite{7,77}. Such Berry phase can be defined as $\phi_{B}=\oint_{C}\mathcal{\mathcal{A}}(\mathbf{k})\cdot d\mathbf{k}$, where $\mathcal{A}(\mathbf{k})=-i\sum_{n\in occ}\left\langle u_{n}(\mathbf{k})\mid\nabla_{k}\mid u_{n}(\mathbf{k})\right\rangle$ is the Berry connection of the occupied state and the closed path \emph{C} threads the NLs. If loop \emph{C} encloses the NL, one has a Berry phase of $\phi_{B}=\pi$; otherwise, $\phi_{B}=0$. We obtain that the Berry phase along a loop \emph{C} threading the NL located on $\hat{M}_{xy}$ (110) or $\hat{\tilde{M}}_{xy}$ ($1\bar{1}0$) mirror plane equals $\pi$. To directly present the intriguing TNL properties in \emph{$\beta$}-PbO$_2$, the projected surface states for the (100) surface are obtained by using the MLWFS, where a flat surface band is clearly visible, as shown in Fig.~\ref{fig4}(a). Interestingly, the drumhead-like surface states are nearly flat with small dispersion, which indicates a high density of states around the Fermi level. In particular, a visual isoenergy band contour of the surface at the Fermi level is shown in Fig.~\ref{fig4}(b), which is one effective approach to observe the drumhead-like surface state. The relevant surface states and isoenergy band contour of the surface in \emph{$\beta$}-PbO$_2$ at (110) surface are obtained in Fig.~\ref{fig4}(c) and Fig.~\ref{fig4}(d). Such drumhead-like surface states of \emph{$\beta$}-PbO$_2$ could be proposed as a route to achieving high-performance catalysis~\cite{78,79}.

To capture the main physics, we develop an effective two-band model based on the symmetry of the 3D NL semimetal \emph{$\beta$}-PbO$_2$. The opposite mirror eigenvalues imply that the corresponding mirror symmetries can be chosen as $\hat{M}_{xy}=-\sigma_{z}$ and $\hat{\tilde{M}}_{xy}=-\sigma_{z}$, and the mirror symmetries are given an additional constraint to Hamiltonian defined as:
\begin{equation}
\begin{split}
\hat{M}_{xy}H(k_{x},k_{y},k_{z})\hat{M}_{xy}^{-1}=H(k_{y},k_{x},k_{z}),\\
\hat{\tilde{M}}_{xy}H(k_{x},k_{y},k_{z})\hat{\tilde{M}}_{xy}^{-1}=H(-k_{y},-k_{x},k_{z}).
\end{split}
\end{equation}
Furthermore, the system contains inversion symmetry ($\mathcal{P}$) and time reversal symmetry ($\mathcal{T}$). The two bands have same parity and thus the inversion operator ($\mathcal{P}$) can be chosen as $\mathcal{P}=\sigma_{0}$. The time reversal symmetry ($\mathcal{T}$) takes the form as $\mathcal{T}=\mathcal{\mathfrak{\mathcal{K}}}$, and the $\mathcal{\mathfrak{\mathcal{K}}}$ is the complex conjugate operator. In addition, the system has the fourfold screw rotation symmetry, and the CBM and VBM along the high-symmetry line $\Gamma \to Z$ can be labeled by the fourfold screw rotation eigenvalues -1 and +1, respectively. Then, the fourfold screw rotation operator can be written as $\tilde{C}_{4z}=-\sigma_{z}$. Here, $\sigma_{0}$ is the identity matrix, and $\sigma_{x,y,z}$ are Pauli matrixes. To characterize the low-energy physics of \emph{$\beta$}-PbO$_2$, the two-band model Hamiltonian under the symmetry constraints can be expressed as:
\begin{equation}
\begin{split}
H(k)=&[a_{0}+a_{1}(k_{x}^{2}+k_{y}^{2})+a_{2}k_{z}^{2}]+[b(k_{x}^{2}-k_{y}^{2})]\sigma_{x}\\
&+[d_{0}+d_{1}(k_{x}^{2}+k_{y}^{2})+d_{2}k_{z}^{2}]\sigma_{z}.
\end{split}
\end{equation}
The corresponding parameters obtained by fitting with the band structure from the HSE06 method are listed in Table \ref{Table2}. The band structure calculated by two-band model Hamiltonian agrees well with the one obtained by the HSE06 method, as shown in Fig.~\ref{fig2}(d).

\begin{table}
  \caption{The parameters for the two-band model Hamiltonian in Eq.(2).}
  \label{Table2}
  \begin{tabular}{c|cccc}
    \hline
    \hline
    \emph{$\beta$}-PbO$_2$	& {\emph{a}$_0$}(eV) &{\emph{a}$_1$}(eV{\AA}$^2$) & {\emph{a}$_2$}(eV{\AA}$^2$)& \emph{b}(eV{\AA}$^2$) \\
    	    & -0.0545	& 24.9 & 65.6 & -328	\\
    \hline
            &{\emph{d}$_0$}(eV) &   {\emph{d}$_1$}(eV{\AA}$^2$) & {\emph{d}$_2$}(eV{\AA}$^2$)& \\
    	    & 0.0605 & -28.5	& -72.4 & \\
    \hline
    \hline
  \end{tabular}
\end{table}

\section{Strain induced phase transition in \emph{$\alpha$}-PbO$_2$ and \emph{$\beta$}-PbO$_2$}

\begin{figure}
\includegraphics[width=1\columnwidth]{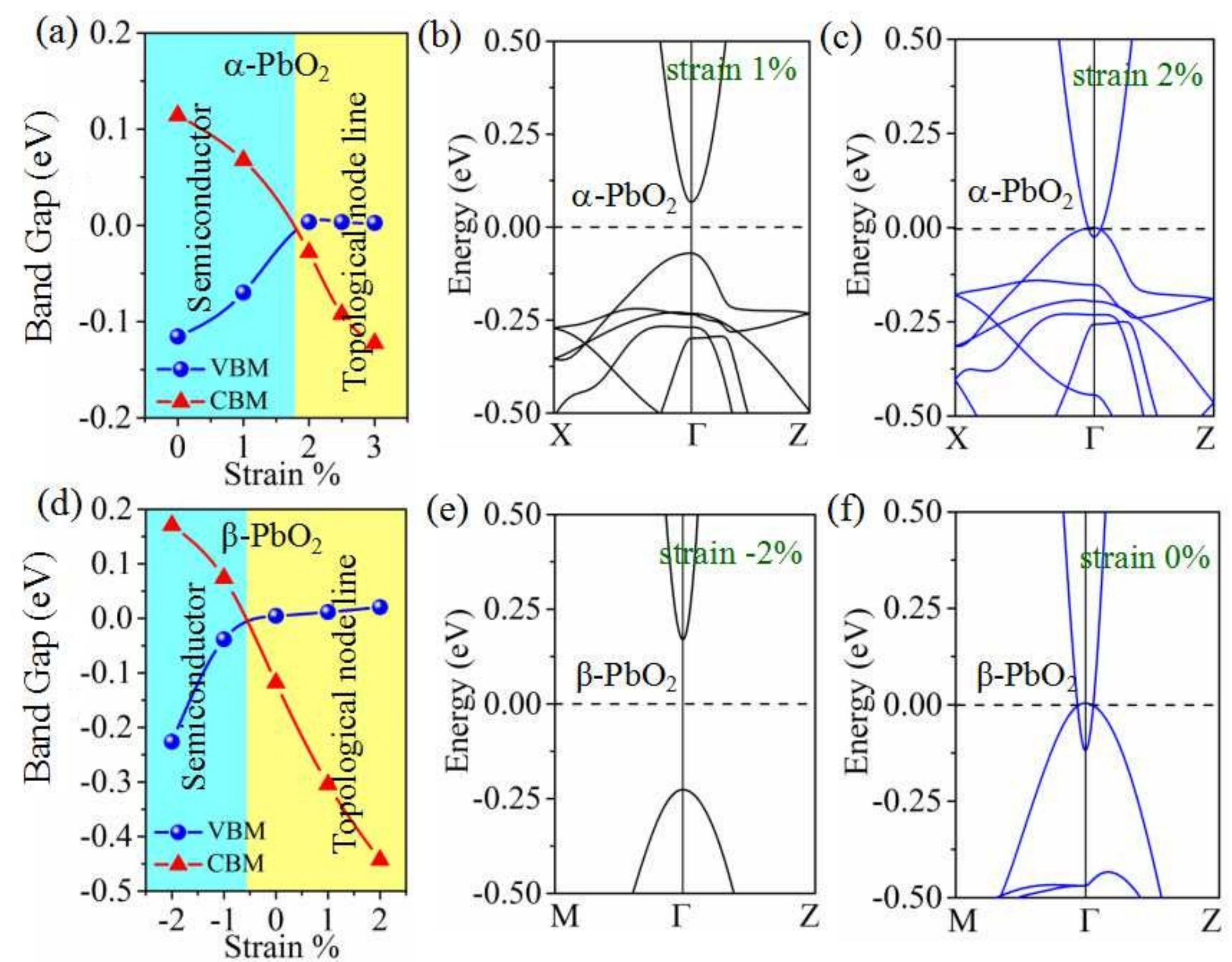}
\caption
{(Color online) Strain-gap relations for (a) \emph{$\alpha$}-PbO$_2$ and (d) \emph{$\beta$}-PbO$_2$ with the different strain. The blue ball is the energy of VBM and the red triangle presents the energy of CBM. Band structures of \emph{$\alpha$}-PbO$_2$ and \emph{$\beta$}-PbO$_2$ are calculated under the hydrostatic strain (b) 1$\% $, (c) 2$\% $, (e) -2$\% $, and (f) 0$\% $, respectively. Here, the light blue zone presents semiconductor and the light yellow pattern denotes TNL.}
\label{fig5}
\end{figure}

On one hand, hydrostatic strain acts a key role in tuning electronic properties in practice. In Fig.~\ref{fig5}(a) and Fig.~\ref{fig5}(d), we plot the evolution of the band gap under different hydrostatic strain via the HSE06 method. The strain-gap curves show that a transition from semiconductor phase to TNL phase in \emph{$\alpha$}-PbO$_2$ under certain tensile strain. At the same time, one can also evaluate the robustness of the NL nature against the hydrostatic strain. Specifically, for \emph{$\beta$}-PbO$_2$, the TNL states are robust for small hydrostatic pressure strain and any hydrostatic tensile strain. For the \emph{$\alpha$}-PbO$_2$, larger hydrostatic tensile strain ($> 1.8\% $) is needed to drive the otherwise semiconductor phase to the TNL states. These valuable findings are very beneficial for the future experimental preparation and make these 3D TSMs highly adaptable to various application environments.

\section{The effect of SOC on the 3D Dirac semimetal state in \emph{$\beta$}-PbO$_2$}

\begin{figure}
\includegraphics[width=1\columnwidth]{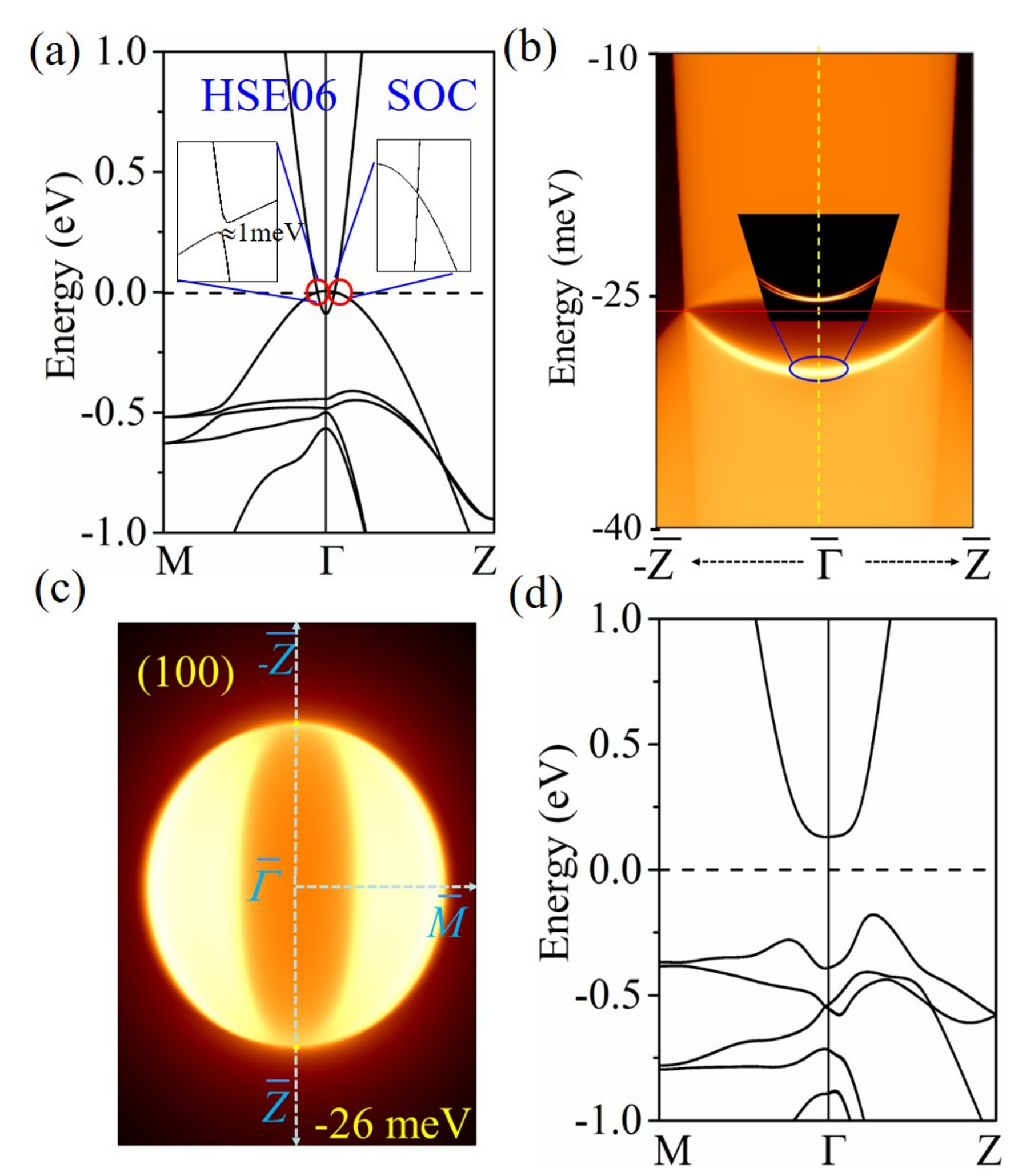}
\caption
{(Color online) (a) The band structures of \emph{$\beta$}-PbO$_2$ via  HSE06 methods with SOC effect. (b) The projected surface states for the (100) surface. (c) The zoom-in isoenergy surface around the $\bar{\Gamma}$ point at the $E_{f}$ =-26 meV for the (100) surface. (d) Band structures of \emph{$\beta$}-PbO$_2$ with SOC under the shear strain by PBE method.}
\label{fig6}
\end{figure}

In the presence of the rather weak SOC effect, strictly speaking, the TNL states in \emph{$\beta$}-PbO$_2$ will evolve into Dirac points along the $\Gamma Z$ line and protected by the fourfold screw rotation ($\tilde{C}_{4z}$) symmetry, but with a tiny gap (~1 meV, much smaller than the room temperature energy scale ~26 meV) opened along the $M \to \Gamma$ path (Fig.~\ref{fig6}(a)). The (100) surface state in Fig.~\ref{fig6}(b) clearly shows the gapless bulk state and a hallmark pair of Fermi arcs along the $-\overline{Z}\rightarrow\overline{\Gamma}\rightarrow\overline{Z}$ path.
To further identify the detailed connection of Fermi arcs, the Fermi arcs at the $E_{f}$ =-26 meV can be specifically shown around these projected Dirac nodes, as shown in Fig.~\ref{fig6}(c). When the $\tilde{C}_{4z}$ is broken, \emph{e.g.}, by the shear strain effect, the band-crossing node will disappear resulting in a trivial band insulator, as shown in Fig.~\ref{fig6}(d).

\section{Topological nodal line states in the derivatives of \emph{$\beta$}-PbO$_2$: \emph{$\beta$}-PbX$_2$ (X=S, Se)}

\begin{figure}
\includegraphics[width=1\columnwidth]{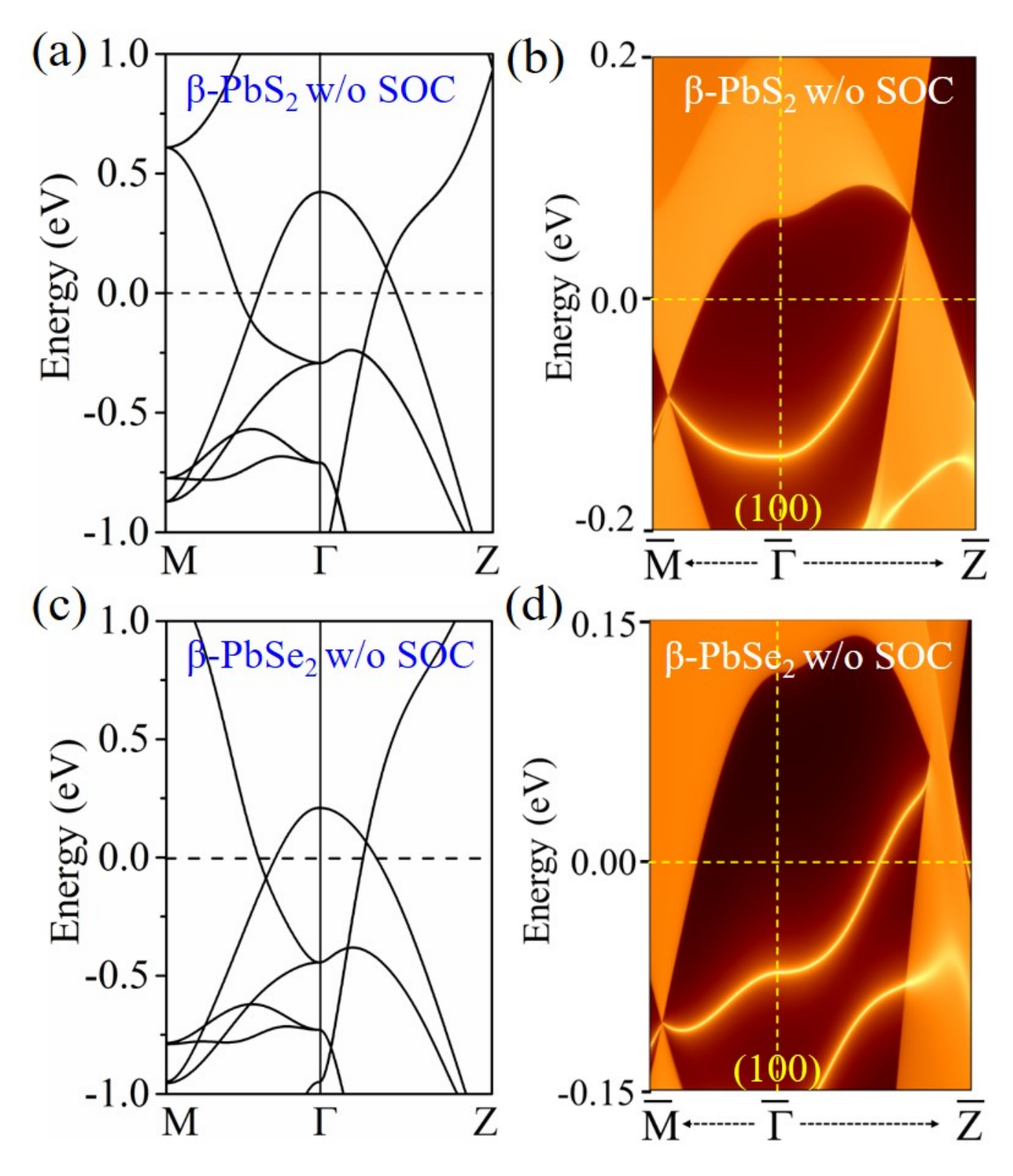}
\caption
{(Color online) In the absence of SOC effect, the band structures of (a) \emph{$\beta$}-PbS$_2$ and (c) \emph{$\beta$}-PbSe$_2$ are calculated via the HSE06 method. (b, d) show the projected surface states for the (100) surface.}
\label{fig7}
\end{figure}

In order to enrich the TNL family, we further explore the electronic properties of their derivatives \emph{$\beta$}-PbX$_2$ (X=S, Se) via the HSE06 method. As twinborn structures of \emph{$\beta$}-PbO$_2$, \emph{i.e.}, \emph{$\beta$}-PbS$_2$ and \emph{$\beta$}-PbSe$_2$, they also show TNL phases, as presented in Fig.~\ref{fig7}(a) and Fig.~\ref{fig7}(c), the band crossing nodes of \emph{$\beta$}-PbS$_2$ and \emph{$\beta$}-PbSe$_2$ are gapless due to the mirror symmetries ($\hat{M}_{xy}$ and  $\hat{\tilde{M}}_{xy}$). The relevant surface state and isoenergy band contour of the surface in \emph{$\beta$}-PbS$_2$ and \emph{$\beta$}-PbSe$_2$ at (100) surface are obtained in Fig.~\ref{fig7}(b) and Fig.~\ref{fig7}(d). These derivatives of \emph{$\beta$}-PbO$_2$ provide more candidates for investigating the novel physics associated with TNL semimetal in the experiment.

\section{Conclusion}

In this paper, based on first-principles calculations and symmetry analysis, two phases of PbO$_2$ (\emph{$\alpha$}-PbO$_2$ and \emph{$\beta$}-PbO$_2$) with stable structure have been discussed. We have reported that \emph{$\beta$}-PbO$_2$ is an intrinsic TNL semimetal due to the mirror symmetry with negligible SOC, and \emph{$\alpha$}-PbO$_2$ has a transition between the semiconductor phase and TNL phase under the hydrostatic strain. Moreover, the drumhead-like surface state of \emph{$\beta$}-PbO$_2$ near the Fermi level is obtained. If the weak SOC is considered, TNL in \emph{$\beta$}-PbO$_2$ evolves into a pair of stable 3D Dirac points as protected by fourfold screw rotation $\tilde{C}_{4z}$ symmetry. In addition, the exploring of \emph{$\beta$}-PbO$_2$ derivatives (\emph{i.e.} \emph{$\beta$}-PbS$_2$ and \emph{$\beta$}-PbSe$_2$) has also been greatly extended the TNL family. Our findings thus provide a valuable playground to explore the intriguing physics in further experiments.

\section{acknowledgments}
This work was supported by the National Natural Science Foundation of China (Grant Nos. 11774028, 11734003, 11574029), the National Key R\&D Program of China (Grant No. 2016YFA0300600), the MOST Project of China (Grant No. 2014CB920903), and Basic Research Funds of Beijing Institute of Technology (Grant No. 2017CX01018).




\bibliography{ref}

\begin{thebibliography}{79}%
\makeatletter
\providecommand \@ifxundefined [1]{%
 \@ifx{#1\undefined}
}%
\providecommand \@ifnum [1]{%
 \ifnum #1\expandafter \@firstoftwo
 \else \expandafter \@secondoftwo
 \fi
}%
\providecommand \@ifx [1]{%
 \ifx #1\expandafter \@firstoftwo
 \else \expandafter \@secondoftwo
 \fi
}%
\providecommand \natexlab [1]{#1}%
\providecommand \enquote  [1]{``#1''}%
\providecommand \bibnamefont  [1]{#1}%
\providecommand \bibfnamefont [1]{#1}%
\providecommand \citenamefont [1]{#1}%
\providecommand \href@noop [0]{\@secondoftwo}%
\providecommand \href [0]{\begingroup \@sanitize@url \@href}%
\providecommand \@href[1]{\@@startlink{#1}\@@href}%
\providecommand \@@href[1]{\endgroup#1\@@endlink}%
\providecommand \@sanitize@url [0]{\catcode `\\12\catcode `\$12\catcode
  `\&12\catcode `\#12\catcode `\^12\catcode `\_12\catcode `\%12\relax}%
\providecommand \@@startlink[1]{}%
\providecommand \@@endlink[0]{}%
\providecommand \url  [0]{\begingroup\@sanitize@url \@url }%
\providecommand \@url [1]{\endgroup\@href {#1}{\urlprefix }}%
\providecommand \urlprefix  [0]{URL }%
\providecommand \Eprint [0]{\href }%
\providecommand \doibase [0]{http://dx.doi.org/}%
\providecommand \selectlanguage [0]{\@gobble}%
\providecommand \bibinfo  [0]{\@secondoftwo}%
\providecommand \bibfield  [0]{\@secondoftwo}%
\providecommand \translation [1]{[#1]}%
\providecommand \BibitemOpen [0]{}%
\providecommand \bibitemStop [0]{}%
\providecommand \bibitemNoStop [0]{.\EOS\space}%
\providecommand \EOS [0]{\spacefactor3000\relax}%
\providecommand \BibitemShut  [1]{\csname bibitem#1\endcsname}%
\let\auto@bib@innerbib\@empty
\bibitem [{\citenamefont {Wan}\ \emph {et~al.}(2011)\citenamefont {Wan},
  \citenamefont {Turner}, \citenamefont {Vishwanath},\ and\ \citenamefont
  {Savrasov}}]{1}%
  \BibitemOpen
  \bibfield  {author} {\bibinfo {author} {\bibfnamefont {X.}~\bibnamefont
  {Wan}}, \bibinfo {author} {\bibfnamefont {A.~M.}\ \bibnamefont {Turner}},
  \bibinfo {author} {\bibfnamefont {A.}~\bibnamefont {Vishwanath}}, \ and\
  \bibinfo {author} {\bibfnamefont {S.~Y.}\ \bibnamefont {Savrasov}},\
  }\href@noop {} {\bibfield  {journal} {\bibinfo  {journal} {Phys. Rev. B}\
  }\textbf {\bibinfo {volume} {83}},\ \bibinfo {pages} {205101} (\bibinfo
  {year} {2011})}\BibitemShut {NoStop}%
\bibitem [{\citenamefont {Burkov}\ \emph {et~al.}(2011)\citenamefont {Burkov},
  \citenamefont {Hook},\ and\ \citenamefont {Balents}}]{2}%
  \BibitemOpen
  \bibfield  {author} {\bibinfo {author} {\bibfnamefont {A.~A.}\ \bibnamefont
  {Burkov}}, \bibinfo {author} {\bibfnamefont {M.~D.}\ \bibnamefont {Hook}}, \
  and\ \bibinfo {author} {\bibfnamefont {L.}~\bibnamefont {Balents}},\
  }\href@noop {} {\bibfield  {journal} {\bibinfo  {journal} {Phys. Rev. B}\
  }\textbf {\bibinfo {volume} {84}},\ \bibinfo {pages} {235126} (\bibinfo
  {year} {2011})}\BibitemShut {NoStop}%
\bibitem [{\citenamefont {Young}\ \emph {et~al.}(2012)\citenamefont {Young},
  \citenamefont {Zaheer}, \citenamefont {Teo}, \citenamefont {Kane},
  \citenamefont {Mele},\ and\ \citenamefont {Rappe}}]{3}%
  \BibitemOpen
  \bibfield  {author} {\bibinfo {author} {\bibfnamefont {S.~M.}\ \bibnamefont
  {Young}}, \bibinfo {author} {\bibfnamefont {S.}~\bibnamefont {Zaheer}},
  \bibinfo {author} {\bibfnamefont {J.~C.~Y.}\ \bibnamefont {Teo}}, \bibinfo
  {author} {\bibfnamefont {C.~L.}\ \bibnamefont {Kane}}, \bibinfo {author}
  {\bibfnamefont {E.~J.}\ \bibnamefont {Mele}}, \ and\ \bibinfo {author}
  {\bibfnamefont {A.~M.}\ \bibnamefont {Rappe}},\ }\href@noop {} {\bibfield
  {journal} {\bibinfo  {journal} {Phys. Rev. Lett.}\ }\textbf {\bibinfo
  {volume} {108}},\ \bibinfo {pages} {140405} (\bibinfo {year}
  {2012})}\BibitemShut {NoStop}%
\bibitem [{\citenamefont {Yang}\ and\ \citenamefont {Nagaosa}(2014)}]{4}%
  \BibitemOpen
  \bibfield  {author} {\bibinfo {author} {\bibfnamefont {B.-J.}\ \bibnamefont
  {Yang}}\ and\ \bibinfo {author} {\bibfnamefont {N.}~\bibnamefont {Nagaosa}},\
  }\href@noop {} {\bibfield  {journal} {\bibinfo  {journal} {Nat. Commun.}\
  }\textbf {\bibinfo {volume} {5}} (\bibinfo {year} {2014})}\BibitemShut
  {NoStop}%
\bibitem [{\citenamefont {Chiu}\ and\ \citenamefont {Schnyder}(2014)}]{5}%
  \BibitemOpen
  \bibfield  {author} {\bibinfo {author} {\bibfnamefont {C.-K.}\ \bibnamefont
  {Chiu}}\ and\ \bibinfo {author} {\bibfnamefont {A.~P.}\ \bibnamefont
  {Schnyder}},\ }\href@noop {} {\bibfield  {journal} {\bibinfo  {journal}
  {Phys. Rev. B}\ }\textbf {\bibinfo {volume} {90}},\ \bibinfo {pages} {205136}
  (\bibinfo {year} {2014})}\BibitemShut {NoStop}%
\bibitem [{\citenamefont {Chen}\ \emph {et~al.}(2018)\citenamefont {Chen},
  \citenamefont {Po}, \citenamefont {Neaton},\ and\ \citenamefont
  {Vishwanath}}]{6}%
  \BibitemOpen
  \bibfield  {author} {\bibinfo {author} {\bibfnamefont {R.}~\bibnamefont
  {Chen}}, \bibinfo {author} {\bibfnamefont {H.~C.}\ \bibnamefont {Po}},
  \bibinfo {author} {\bibfnamefont {J.~B.}\ \bibnamefont {Neaton}}, \ and\
  \bibinfo {author} {\bibfnamefont {A.}~\bibnamefont {Vishwanath}},\
  }\href@noop {} {\bibfield  {journal} {\bibinfo  {journal} {Nat. Phys.}\
  }\textbf {\bibinfo {volume} {14}},\ \bibinfo {pages} {55} (\bibinfo {year}
  {2018})}\BibitemShut {NoStop}%
\bibitem [{\citenamefont {Fang}\ \emph {et~al.}(2016)\citenamefont {Fang},
  \citenamefont {Weng}, \citenamefont {Dai},\ and\ \citenamefont {Fang}}]{7}%
  \BibitemOpen
  \bibfield  {author} {\bibinfo {author} {\bibfnamefont {C.}~\bibnamefont
  {Fang}}, \bibinfo {author} {\bibfnamefont {H.}~\bibnamefont {Weng}}, \bibinfo
  {author} {\bibfnamefont {X.}~\bibnamefont {Dai}}, \ and\ \bibinfo {author}
  {\bibfnamefont {Z.}~\bibnamefont {Fang}},\ }\href@noop {} {\bibfield
  {journal} {\bibinfo  {journal} {Chin. Phys. B}\ }\textbf {\bibinfo {volume}
  {25}},\ \bibinfo {pages} {117106} (\bibinfo {year} {2016})}\BibitemShut
  {NoStop}%
\bibitem [{\citenamefont {Huang}\ \emph {et~al.}(2017)\citenamefont {Huang},
  \citenamefont {Jin},\ and\ \citenamefont {Liu}}]{8}%
  \BibitemOpen
  \bibfield  {author} {\bibinfo {author} {\bibfnamefont {H.}~\bibnamefont
  {Huang}}, \bibinfo {author} {\bibfnamefont {K.-H.}\ \bibnamefont {Jin}}, \
  and\ \bibinfo {author} {\bibfnamefont {F.}~\bibnamefont {Liu}},\ }\href@noop
  {} {\bibfield  {journal} {\bibinfo  {journal} {Phys. Rev. B}\ }\textbf
  {\bibinfo {volume} {96}},\ \bibinfo {pages} {115106} (\bibinfo {year}
  {2017})}\BibitemShut {NoStop}%
\bibitem [{\citenamefont {Wang}\ \emph {et~al.}(2012)\citenamefont {Wang},
  \citenamefont {Sun}, \citenamefont {Chen}, \citenamefont {Franchini},
  \citenamefont {Xu}, \citenamefont {Weng}, \citenamefont {Dai},\ and\
  \citenamefont {Fang}}]{9}%
  \BibitemOpen
  \bibfield  {author} {\bibinfo {author} {\bibfnamefont {Z.}~\bibnamefont
  {Wang}}, \bibinfo {author} {\bibfnamefont {Y.}~\bibnamefont {Sun}}, \bibinfo
  {author} {\bibfnamefont {X.-Q.}\ \bibnamefont {Chen}}, \bibinfo {author}
  {\bibfnamefont {C.}~\bibnamefont {Franchini}}, \bibinfo {author}
  {\bibfnamefont {G.}~\bibnamefont {Xu}}, \bibinfo {author} {\bibfnamefont
  {H.}~\bibnamefont {Weng}}, \bibinfo {author} {\bibfnamefont {X.}~\bibnamefont
  {Dai}}, \ and\ \bibinfo {author} {\bibfnamefont {Z.}~\bibnamefont {Fang}},\
  }\href@noop {} {\bibfield  {journal} {\bibinfo  {journal} {Phys. Rev. B}\
  }\textbf {\bibinfo {volume} {85}},\ \bibinfo {pages} {195320} (\bibinfo
  {year} {2012})}\BibitemShut {NoStop}%
\bibitem [{\citenamefont {Wang}\ \emph {et~al.}(2013)\citenamefont {Wang},
  \citenamefont {Weng}, \citenamefont {Wu}, \citenamefont {Dai},\ and\
  \citenamefont {Fang}}]{10}%
  \BibitemOpen
  \bibfield  {author} {\bibinfo {author} {\bibfnamefont {Z.}~\bibnamefont
  {Wang}}, \bibinfo {author} {\bibfnamefont {H.}~\bibnamefont {Weng}}, \bibinfo
  {author} {\bibfnamefont {Q.}~\bibnamefont {Wu}}, \bibinfo {author}
  {\bibfnamefont {X.}~\bibnamefont {Dai}}, \ and\ \bibinfo {author}
  {\bibfnamefont {Z.}~\bibnamefont {Fang}},\ }\href@noop {} {\bibfield
  {journal} {\bibinfo  {journal} {Phys. Rev. B}\ }\textbf {\bibinfo {volume}
  {88}},\ \bibinfo {pages} {125427} (\bibinfo {year} {2013})}\BibitemShut
  {NoStop}%
\bibitem [{\citenamefont {Huang}\ \emph
  {et~al.}(2016{\natexlab{a}})\citenamefont {Huang}, \citenamefont {Zhou},\
  and\ \citenamefont {Duan}}]{11}%
  \BibitemOpen
  \bibfield  {author} {\bibinfo {author} {\bibfnamefont {H.}~\bibnamefont
  {Huang}}, \bibinfo {author} {\bibfnamefont {S.}~\bibnamefont {Zhou}}, \ and\
  \bibinfo {author} {\bibfnamefont {W.}~\bibnamefont {Duan}},\ }\href@noop {}
  {\bibfield  {journal} {\bibinfo  {journal} {Phys. Rev. B}\ }\textbf {\bibinfo
  {volume} {94}},\ \bibinfo {pages} {121117} (\bibinfo {year}
  {2016}{\natexlab{a}})}\BibitemShut {NoStop}%
\bibitem [{\citenamefont {Chang}\ \emph {et~al.}(2017)\citenamefont {Chang},
  \citenamefont {Xu}, \citenamefont {Sanchez}, \citenamefont {Tsai},
  \citenamefont {Huang}, \citenamefont {Chang}, \citenamefont {Hsu},
  \citenamefont {Bian}, \citenamefont {Belopolski}, \citenamefont {Yu},
  \citenamefont {Yang}, \citenamefont {Neupert}, \citenamefont {Jeng},
  \citenamefont {Lin},\ and\ \citenamefont {Hasan}}]{12}%
  \BibitemOpen
  \bibfield  {author} {\bibinfo {author} {\bibfnamefont {T.-R.}\ \bibnamefont
  {Chang}}, \bibinfo {author} {\bibfnamefont {S.-Y.}\ \bibnamefont {Xu}},
  \bibinfo {author} {\bibfnamefont {D.~S.}\ \bibnamefont {Sanchez}}, \bibinfo
  {author} {\bibfnamefont {W.-F.}\ \bibnamefont {Tsai}}, \bibinfo {author}
  {\bibfnamefont {S.-M.}\ \bibnamefont {Huang}}, \bibinfo {author}
  {\bibfnamefont {G.}~\bibnamefont {Chang}}, \bibinfo {author} {\bibfnamefont
  {C.-H.}\ \bibnamefont {Hsu}}, \bibinfo {author} {\bibfnamefont
  {G.}~\bibnamefont {Bian}}, \bibinfo {author} {\bibfnamefont {I.}~\bibnamefont
  {Belopolski}}, \bibinfo {author} {\bibfnamefont {Z.-M.}\ \bibnamefont {Yu}},
  \bibinfo {author} {\bibfnamefont {S.~A.}\ \bibnamefont {Yang}}, \bibinfo
  {author} {\bibfnamefont {T.}~\bibnamefont {Neupert}}, \bibinfo {author}
  {\bibfnamefont {H.-T.}\ \bibnamefont {Jeng}}, \bibinfo {author}
  {\bibfnamefont {H.}~\bibnamefont {Lin}}, \ and\ \bibinfo {author}
  {\bibfnamefont {M.~Z.}\ \bibnamefont {Hasan}},\ }\href@noop {} {\bibfield
  {journal} {\bibinfo  {journal} {Phys. Rev. Lett.}\ }\textbf {\bibinfo
  {volume} {119}},\ \bibinfo {pages} {026404} (\bibinfo {year}
  {2017})}\BibitemShut {NoStop}%
\bibitem [{\citenamefont {Cao}\ \emph {et~al.}(2017)\citenamefont {Cao},
  \citenamefont {Tang}, \citenamefont {Xu}, \citenamefont {Wu}, \citenamefont
  {Gu},\ and\ \citenamefont {Duan}}]{13}%
  \BibitemOpen
  \bibfield  {author} {\bibinfo {author} {\bibfnamefont {W.}~\bibnamefont
  {Cao}}, \bibinfo {author} {\bibfnamefont {P.}~\bibnamefont {Tang}}, \bibinfo
  {author} {\bibfnamefont {Y.}~\bibnamefont {Xu}}, \bibinfo {author}
  {\bibfnamefont {J.}~\bibnamefont {Wu}}, \bibinfo {author} {\bibfnamefont
  {B.-L.}\ \bibnamefont {Gu}}, \ and\ \bibinfo {author} {\bibfnamefont
  {W.}~\bibnamefont {Duan}},\ }\href@noop {} {\bibfield  {journal} {\bibinfo
  {journal} {Phys. Rev. B}\ }\textbf {\bibinfo {volume} {96}},\ \bibinfo
  {pages} {115203} (\bibinfo {year} {2017})}\BibitemShut {NoStop}%
\bibitem [{\citenamefont {Xu}\ \emph {et~al.}(2011)\citenamefont {Xu},
  \citenamefont {Weng}, \citenamefont {Wang}, \citenamefont {Dai},\ and\
  \citenamefont {Fang}}]{14}%
  \BibitemOpen
  \bibfield  {author} {\bibinfo {author} {\bibfnamefont {G.}~\bibnamefont
  {Xu}}, \bibinfo {author} {\bibfnamefont {H.}~\bibnamefont {Weng}}, \bibinfo
  {author} {\bibfnamefont {Z.}~\bibnamefont {Wang}}, \bibinfo {author}
  {\bibfnamefont {X.}~\bibnamefont {Dai}}, \ and\ \bibinfo {author}
  {\bibfnamefont {Z.}~\bibnamefont {Fang}},\ }\href@noop {} {\bibfield
  {journal} {\bibinfo  {journal} {Phys. Rev. Lett.}\ }\textbf {\bibinfo
  {volume} {107}},\ \bibinfo {pages} {186806} (\bibinfo {year}
  {2011})}\BibitemShut {NoStop}%
\bibitem [{\citenamefont {Weng}\ \emph
  {et~al.}(2015{\natexlab{a}})\citenamefont {Weng}, \citenamefont {Fang},
  \citenamefont {Fang}, \citenamefont {Bernevig},\ and\ \citenamefont
  {Dai}}]{15}%
  \BibitemOpen
  \bibfield  {author} {\bibinfo {author} {\bibfnamefont {H.}~\bibnamefont
  {Weng}}, \bibinfo {author} {\bibfnamefont {C.}~\bibnamefont {Fang}}, \bibinfo
  {author} {\bibfnamefont {Z.}~\bibnamefont {Fang}}, \bibinfo {author}
  {\bibfnamefont {B.~A.}\ \bibnamefont {Bernevig}}, \ and\ \bibinfo {author}
  {\bibfnamefont {X.}~\bibnamefont {Dai}},\ }\href@noop {} {\bibfield
  {journal} {\bibinfo  {journal} {Phys. Rev. X}\ }\textbf {\bibinfo {volume}
  {5}},\ \bibinfo {pages} {011029} (\bibinfo {year}
  {2015}{\natexlab{a}})}\BibitemShut {NoStop}%
\bibitem [{\citenamefont {Sun}\ \emph {et~al.}(2015)\citenamefont {Sun},
  \citenamefont {Wu}, \citenamefont {Ali}, \citenamefont {Felser},\ and\
  \citenamefont {Yan}}]{16}%
  \BibitemOpen
  \bibfield  {author} {\bibinfo {author} {\bibfnamefont {Y.}~\bibnamefont
  {Sun}}, \bibinfo {author} {\bibfnamefont {S.-C.}\ \bibnamefont {Wu}},
  \bibinfo {author} {\bibfnamefont {M.~N.}\ \bibnamefont {Ali}}, \bibinfo
  {author} {\bibfnamefont {C.}~\bibnamefont {Felser}}, \ and\ \bibinfo {author}
  {\bibfnamefont {B.}~\bibnamefont {Yan}},\ }\href@noop {} {\bibfield
  {journal} {\bibinfo  {journal} {Phys. Rev. B}\ }\textbf {\bibinfo {volume}
  {92}},\ \bibinfo {pages} {161107} (\bibinfo {year} {2015})}\BibitemShut
  {NoStop}%
\bibitem [{\citenamefont {Soluyanov}\ \emph {et~al.}(2015)\citenamefont
  {Soluyanov}, \citenamefont {Gresch}, \citenamefont {Wang}, \citenamefont
  {Wu}, \citenamefont {Troyer}, \citenamefont {Dai},\ and\ \citenamefont
  {Bernevig}}]{17}%
  \BibitemOpen
  \bibfield  {author} {\bibinfo {author} {\bibfnamefont {A.~A.}\ \bibnamefont
  {Soluyanov}}, \bibinfo {author} {\bibfnamefont {D.}~\bibnamefont {Gresch}},
  \bibinfo {author} {\bibfnamefont {Z.}~\bibnamefont {Wang}}, \bibinfo {author}
  {\bibfnamefont {Q.}~\bibnamefont {Wu}}, \bibinfo {author} {\bibfnamefont
  {M.}~\bibnamefont {Troyer}}, \bibinfo {author} {\bibfnamefont
  {X.}~\bibnamefont {Dai}}, \ and\ \bibinfo {author} {\bibfnamefont {B.~A.}\
  \bibnamefont {Bernevig}},\ }\href@noop {} {\bibfield  {journal} {\bibinfo
  {journal} {Nature}\ }\textbf {\bibinfo {volume} {527}},\ \bibinfo {pages}
  {495} (\bibinfo {year} {2015})}\BibitemShut {NoStop}%
\bibitem [{\citenamefont {Xu}\ \emph {et~al.}(2015)\citenamefont {Xu},
  \citenamefont {Belopolski}, \citenamefont {Alidoust}, \citenamefont
  {Neupane}, \citenamefont {Bian}, \citenamefont {Zhang}, \citenamefont
  {Sankar}, \citenamefont {Chang}, \citenamefont {Yuan}, \citenamefont {Lee},
  \citenamefont {Huang}, \citenamefont {Zheng}, \citenamefont {Ma},
  \citenamefont {Sanchez}, \citenamefont {Wang}, \citenamefont {Bansil},
  \citenamefont {Chou}, \citenamefont {Shibayev}, \citenamefont {Lin},
  \citenamefont {Jia},\ and\ \citenamefont {Hasan}}]{18}%
  \BibitemOpen
  \bibfield  {author} {\bibinfo {author} {\bibfnamefont {S.-Y.}\ \bibnamefont
  {Xu}}, \bibinfo {author} {\bibfnamefont {I.}~\bibnamefont {Belopolski}},
  \bibinfo {author} {\bibfnamefont {N.}~\bibnamefont {Alidoust}}, \bibinfo
  {author} {\bibfnamefont {M.}~\bibnamefont {Neupane}}, \bibinfo {author}
  {\bibfnamefont {G.}~\bibnamefont {Bian}}, \bibinfo {author} {\bibfnamefont
  {C.}~\bibnamefont {Zhang}}, \bibinfo {author} {\bibfnamefont
  {R.}~\bibnamefont {Sankar}}, \bibinfo {author} {\bibfnamefont
  {G.}~\bibnamefont {Chang}}, \bibinfo {author} {\bibfnamefont
  {Z.}~\bibnamefont {Yuan}}, \bibinfo {author} {\bibfnamefont {C.-C.}\
  \bibnamefont {Lee}}, \bibinfo {author} {\bibfnamefont {S.-M.}\ \bibnamefont
  {Huang}}, \bibinfo {author} {\bibfnamefont {H.}~\bibnamefont {Zheng}},
  \bibinfo {author} {\bibfnamefont {J.}~\bibnamefont {Ma}}, \bibinfo {author}
  {\bibfnamefont {D.~S.}\ \bibnamefont {Sanchez}}, \bibinfo {author}
  {\bibfnamefont {B.}~\bibnamefont {Wang}}, \bibinfo {author} {\bibfnamefont
  {A.}~\bibnamefont {Bansil}}, \bibinfo {author} {\bibfnamefont
  {F.}~\bibnamefont {Chou}}, \bibinfo {author} {\bibfnamefont {P.~P.}\
  \bibnamefont {Shibayev}}, \bibinfo {author} {\bibfnamefont {H.}~\bibnamefont
  {Lin}}, \bibinfo {author} {\bibfnamefont {S.}~\bibnamefont {Jia}}, \ and\
  \bibinfo {author} {\bibfnamefont {M.~Z.}\ \bibnamefont {Hasan}},\ }\href@noop
  {} {\bibfield  {journal} {\bibinfo  {journal} {Science}\ }\textbf {\bibinfo
  {volume} {349}},\ \bibinfo {pages} {613} (\bibinfo {year}
  {2015})}\BibitemShut {NoStop}%
\bibitem [{\citenamefont {Chang}\ \emph {et~al.}(2016)\citenamefont {Chang},
  \citenamefont {Xu}, \citenamefont {Sanchez}, \citenamefont {Huang},
  \citenamefont {Lee}, \citenamefont {Chang}, \citenamefont {Bian},
  \citenamefont {Zheng}, \citenamefont {Belopolski}, \citenamefont {Alidoust},
  \citenamefont {Jeng}, \citenamefont {Bansil}, \citenamefont {Lin},\ and\
  \citenamefont {Hasan}}]{19}%
  \BibitemOpen
  \bibfield  {author} {\bibinfo {author} {\bibfnamefont {G.}~\bibnamefont
  {Chang}}, \bibinfo {author} {\bibfnamefont {S.-Y.}\ \bibnamefont {Xu}},
  \bibinfo {author} {\bibfnamefont {D.~S.}\ \bibnamefont {Sanchez}}, \bibinfo
  {author} {\bibfnamefont {S.-M.}\ \bibnamefont {Huang}}, \bibinfo {author}
  {\bibfnamefont {C.-C.}\ \bibnamefont {Lee}}, \bibinfo {author} {\bibfnamefont
  {T.-R.}\ \bibnamefont {Chang}}, \bibinfo {author} {\bibfnamefont
  {G.}~\bibnamefont {Bian}}, \bibinfo {author} {\bibfnamefont {H.}~\bibnamefont
  {Zheng}}, \bibinfo {author} {\bibfnamefont {I.}~\bibnamefont {Belopolski}},
  \bibinfo {author} {\bibfnamefont {N.}~\bibnamefont {Alidoust}}, \bibinfo
  {author} {\bibfnamefont {H.-T.}\ \bibnamefont {Jeng}}, \bibinfo {author}
  {\bibfnamefont {A.}~\bibnamefont {Bansil}}, \bibinfo {author} {\bibfnamefont
  {H.}~\bibnamefont {Lin}}, \ and\ \bibinfo {author} {\bibfnamefont {M.~Z.}\
  \bibnamefont {Hasan}},\ }\href@noop {} {\bibfield  {journal} {\bibinfo
  {journal} {Sci. Adv.}\ }\textbf {\bibinfo {volume} {2}} (\bibinfo {year}
  {2016})}\BibitemShut {NoStop}%
\bibitem [{\citenamefont {Ruan}\ \emph {et~al.}(2016)\citenamefont {Ruan},
  \citenamefont {Jian}, \citenamefont {Zhang}, \citenamefont {Yao},
  \citenamefont {Zhang}, \citenamefont {Zhang},\ and\ \citenamefont
  {Xing}}]{20}%
  \BibitemOpen
  \bibfield  {author} {\bibinfo {author} {\bibfnamefont {J.}~\bibnamefont
  {Ruan}}, \bibinfo {author} {\bibfnamefont {S.-K.}\ \bibnamefont {Jian}},
  \bibinfo {author} {\bibfnamefont {D.}~\bibnamefont {Zhang}}, \bibinfo
  {author} {\bibfnamefont {H.}~\bibnamefont {Yao}}, \bibinfo {author}
  {\bibfnamefont {H.}~\bibnamefont {Zhang}}, \bibinfo {author} {\bibfnamefont
  {S.-C.}\ \bibnamefont {Zhang}}, \ and\ \bibinfo {author} {\bibfnamefont
  {D.}~\bibnamefont {Xing}},\ }\href@noop {} {\bibfield  {journal} {\bibinfo
  {journal} {Phys. Rev. Lett.}\ }\textbf {\bibinfo {volume} {116}},\ \bibinfo
  {pages} {226801} (\bibinfo {year} {2016})}\BibitemShut {NoStop}%
\bibitem [{\citenamefont {Aut\`es}\ \emph {et~al.}(2016)\citenamefont
  {Aut\`es}, \citenamefont {Gresch}, \citenamefont {Troyer}, \citenamefont
  {Soluyanov},\ and\ \citenamefont {Yazyev}}]{21}%
  \BibitemOpen
  \bibfield  {author} {\bibinfo {author} {\bibfnamefont {G.}~\bibnamefont
  {Aut\`es}}, \bibinfo {author} {\bibfnamefont {D.}~\bibnamefont {Gresch}},
  \bibinfo {author} {\bibfnamefont {M.}~\bibnamefont {Troyer}}, \bibinfo
  {author} {\bibfnamefont {A.~A.}\ \bibnamefont {Soluyanov}}, \ and\ \bibinfo
  {author} {\bibfnamefont {O.~V.}\ \bibnamefont {Yazyev}},\ }\href@noop {}
  {\bibfield  {journal} {\bibinfo  {journal} {Phys. Rev. Lett.}\ }\textbf
  {\bibinfo {volume} {117}},\ \bibinfo {pages} {066402} (\bibinfo {year}
  {2016})}\BibitemShut {NoStop}%
\bibitem [{\citenamefont {Yu}\ \emph {et~al.}(2016)\citenamefont {Yu},
  \citenamefont {Yao},\ and\ \citenamefont {Yang}}]{22}%
  \BibitemOpen
  \bibfield  {author} {\bibinfo {author} {\bibfnamefont {Z.-M.}\ \bibnamefont
  {Yu}}, \bibinfo {author} {\bibfnamefont {Y.}~\bibnamefont {Yao}}, \ and\
  \bibinfo {author} {\bibfnamefont {S.~A.}\ \bibnamefont {Yang}},\ }\href@noop
  {} {\bibfield  {journal} {\bibinfo  {journal} {Phys. Rev. Lett.}\ }\textbf
  {\bibinfo {volume} {117}},\ \bibinfo {pages} {077202} (\bibinfo {year}
  {2016})}\BibitemShut {NoStop}%
\bibitem [{\citenamefont {Liu}\ \emph {et~al.}(2016{\natexlab{a}})\citenamefont
  {Liu}, \citenamefont {Zhou}, \citenamefont {Yao},\ and\ \citenamefont
  {Zhang}}]{23}%
  \BibitemOpen
  \bibfield  {author} {\bibinfo {author} {\bibfnamefont {C.-C.}\ \bibnamefont
  {Liu}}, \bibinfo {author} {\bibfnamefont {J.-J.}\ \bibnamefont {Zhou}},
  \bibinfo {author} {\bibfnamefont {Y.}~\bibnamefont {Yao}}, \ and\ \bibinfo
  {author} {\bibfnamefont {F.}~\bibnamefont {Zhang}},\ }\href@noop {}
  {\bibfield  {journal} {\bibinfo  {journal} {Phys. Rev. Lett.}\ }\textbf
  {\bibinfo {volume} {116}},\ \bibinfo {pages} {066801} (\bibinfo {year}
  {2016}{\natexlab{a}})}\BibitemShut {NoStop}%
\bibitem [{\citenamefont {Zhang}\ \emph {et~al.}(2016)\citenamefont {Zhang},
  \citenamefont {Lu},\ and\ \citenamefont {Shen}}]{24}%
  \BibitemOpen
  \bibfield  {author} {\bibinfo {author} {\bibfnamefont {S.-B.}\ \bibnamefont
  {Zhang}}, \bibinfo {author} {\bibfnamefont {H.-Z.}\ \bibnamefont {Lu}}, \
  and\ \bibinfo {author} {\bibfnamefont {S.-Q.}\ \bibnamefont {Shen}},\
  }\href@noop {} {\bibfield  {journal} {\bibinfo  {journal} {New J. Phys.}\
  }\textbf {\bibinfo {volume} {18}},\ \bibinfo {pages} {053039} (\bibinfo
  {year} {2016})}\BibitemShut {NoStop}%
\bibitem [{\citenamefont {Mullen}\ \emph {et~al.}(2015)\citenamefont {Mullen},
  \citenamefont {Uchoa},\ and\ \citenamefont {Glatzhofer}}]{25}%
  \BibitemOpen
  \bibfield  {author} {\bibinfo {author} {\bibfnamefont {K.}~\bibnamefont
  {Mullen}}, \bibinfo {author} {\bibfnamefont {B.}~\bibnamefont {Uchoa}}, \
  and\ \bibinfo {author} {\bibfnamefont {D.~T.}\ \bibnamefont {Glatzhofer}},\
  }\href@noop {} {\bibfield  {journal} {\bibinfo  {journal} {Phys. Rev. Lett.}\
  }\textbf {\bibinfo {volume} {115}},\ \bibinfo {pages} {026403} (\bibinfo
  {year} {2015})}\BibitemShut {NoStop}%
\bibitem [{\citenamefont {Fang}\ \emph {et~al.}(2015)\citenamefont {Fang},
  \citenamefont {Chen}, \citenamefont {Kee},\ and\ \citenamefont {Fu}}]{26}%
  \BibitemOpen
  \bibfield  {author} {\bibinfo {author} {\bibfnamefont {C.}~\bibnamefont
  {Fang}}, \bibinfo {author} {\bibfnamefont {Y.}~\bibnamefont {Chen}}, \bibinfo
  {author} {\bibfnamefont {H.-Y.}\ \bibnamefont {Kee}}, \ and\ \bibinfo
  {author} {\bibfnamefont {L.}~\bibnamefont {Fu}},\ }\href@noop {} {\bibfield
  {journal} {\bibinfo  {journal} {Phys. Rev. B}\ }\textbf {\bibinfo {volume}
  {92}},\ \bibinfo {pages} {081201} (\bibinfo {year} {2015})}\BibitemShut
  {NoStop}%
\bibitem [{\citenamefont {Weng}\ \emph
  {et~al.}(2015{\natexlab{b}})\citenamefont {Weng}, \citenamefont {Liang},
  \citenamefont {Xu}, \citenamefont {Yu}, \citenamefont {Fang}, \citenamefont
  {Dai},\ and\ \citenamefont {Kawazoe}}]{27}%
  \BibitemOpen
  \bibfield  {author} {\bibinfo {author} {\bibfnamefont {H.}~\bibnamefont
  {Weng}}, \bibinfo {author} {\bibfnamefont {Y.}~\bibnamefont {Liang}},
  \bibinfo {author} {\bibfnamefont {Q.}~\bibnamefont {Xu}}, \bibinfo {author}
  {\bibfnamefont {R.}~\bibnamefont {Yu}}, \bibinfo {author} {\bibfnamefont
  {Z.}~\bibnamefont {Fang}}, \bibinfo {author} {\bibfnamefont {X.}~\bibnamefont
  {Dai}}, \ and\ \bibinfo {author} {\bibfnamefont {Y.}~\bibnamefont
  {Kawazoe}},\ }\href@noop {} {\bibfield  {journal} {\bibinfo  {journal} {Phys.
  Rev. B}\ }\textbf {\bibinfo {volume} {92}},\ \bibinfo {pages} {045108}
  (\bibinfo {year} {2015}{\natexlab{b}})}\BibitemShut {NoStop}%
\bibitem [{\citenamefont {Kim}\ \emph {et~al.}(2015{\natexlab{a}})\citenamefont
  {Kim}, \citenamefont {Wieder}, \citenamefont {Kane},\ and\ \citenamefont
  {Rappe}}]{28}%
  \BibitemOpen
  \bibfield  {author} {\bibinfo {author} {\bibfnamefont {Y.}~\bibnamefont
  {Kim}}, \bibinfo {author} {\bibfnamefont {B.~J.}\ \bibnamefont {Wieder}},
  \bibinfo {author} {\bibfnamefont {C.~L.}\ \bibnamefont {Kane}}, \ and\
  \bibinfo {author} {\bibfnamefont {A.~M.}\ \bibnamefont {Rappe}},\ }\href@noop
  {} {\bibfield  {journal} {\bibinfo  {journal} {Phys. Rev. Lett.}\ }\textbf
  {\bibinfo {volume} {115}},\ \bibinfo {pages} {036806} (\bibinfo {year}
  {2015}{\natexlab{a}})}\BibitemShut {NoStop}%
\bibitem [{\citenamefont {Ezawa}(2016)}]{29}%
  \BibitemOpen
  \bibfield  {author} {\bibinfo {author} {\bibfnamefont {M.}~\bibnamefont
  {Ezawa}},\ }\href@noop {} {\bibfield  {journal} {\bibinfo  {journal} {Phys.
  Rev. Lett.}\ }\textbf {\bibinfo {volume} {116}},\ \bibinfo {pages} {127202}
  (\bibinfo {year} {2016})}\BibitemShut {NoStop}%
\bibitem [{\citenamefont {Hu}\ \emph {et~al.}(2016)\citenamefont {Hu},
  \citenamefont {Tang}, \citenamefont {Liu}, \citenamefont {Liu}, \citenamefont
  {Zhu}, \citenamefont {Graf}, \citenamefont {Myhro}, \citenamefont {Tran},
  \citenamefont {Lau}, \citenamefont {Wei} \emph {et~al.}}]{30}%
  \BibitemOpen
  \bibfield  {author} {\bibinfo {author} {\bibfnamefont {J.}~\bibnamefont
  {Hu}}, \bibinfo {author} {\bibfnamefont {Z.}~\bibnamefont {Tang}}, \bibinfo
  {author} {\bibfnamefont {J.}~\bibnamefont {Liu}}, \bibinfo {author}
  {\bibfnamefont {X.}~\bibnamefont {Liu}}, \bibinfo {author} {\bibfnamefont
  {Y.}~\bibnamefont {Zhu}}, \bibinfo {author} {\bibfnamefont {D.}~\bibnamefont
  {Graf}}, \bibinfo {author} {\bibfnamefont {K.}~\bibnamefont {Myhro}},
  \bibinfo {author} {\bibfnamefont {S.}~\bibnamefont {Tran}}, \bibinfo {author}
  {\bibfnamefont {C.~N.}\ \bibnamefont {Lau}}, \bibinfo {author} {\bibfnamefont
  {J.}~\bibnamefont {Wei}},  \emph {et~al.},\ }\href@noop {} {\bibfield
  {journal} {\bibinfo  {journal} {Phys. Rev. Lett.}\ }\textbf {\bibinfo
  {volume} {117}},\ \bibinfo {pages} {016602} (\bibinfo {year}
  {2016})}\BibitemShut {NoStop}%
\bibitem [{\citenamefont {Bian}\ \emph {et~al.}(2016)\citenamefont {Bian},
  \citenamefont {Chang}, \citenamefont {Sankar}, \citenamefont {Xu},
  \citenamefont {Zheng}, \citenamefont {Neupert}, \citenamefont {Chiu},
  \citenamefont {Huang}, \citenamefont {Chang}, \citenamefont {Belopolski}
  \emph {et~al.}}]{31}%
  \BibitemOpen
  \bibfield  {author} {\bibinfo {author} {\bibfnamefont {G.}~\bibnamefont
  {Bian}}, \bibinfo {author} {\bibfnamefont {T.-R.}\ \bibnamefont {Chang}},
  \bibinfo {author} {\bibfnamefont {R.}~\bibnamefont {Sankar}}, \bibinfo
  {author} {\bibfnamefont {S.-Y.}\ \bibnamefont {Xu}}, \bibinfo {author}
  {\bibfnamefont {H.}~\bibnamefont {Zheng}}, \bibinfo {author} {\bibfnamefont
  {T.}~\bibnamefont {Neupert}}, \bibinfo {author} {\bibfnamefont {C.-K.}\
  \bibnamefont {Chiu}}, \bibinfo {author} {\bibfnamefont {S.-M.}\ \bibnamefont
  {Huang}}, \bibinfo {author} {\bibfnamefont {G.}~\bibnamefont {Chang}},
  \bibinfo {author} {\bibfnamefont {I.}~\bibnamefont {Belopolski}},  \emph
  {et~al.},\ }\href@noop {} {\bibfield  {journal} {\bibinfo  {journal} {Nat.
  Commun.}\ }\textbf {\bibinfo {volume} {7}},\ \bibinfo {pages} {10556}
  (\bibinfo {year} {2016})}\BibitemShut {NoStop}%
\bibitem [{\citenamefont {Chan}\ \emph {et~al.}(2016)\citenamefont {Chan},
  \citenamefont {Chiu}, \citenamefont {Chou},\ and\ \citenamefont
  {Schnyder}}]{32}%
  \BibitemOpen
  \bibfield  {author} {\bibinfo {author} {\bibfnamefont {Y.-H.}\ \bibnamefont
  {Chan}}, \bibinfo {author} {\bibfnamefont {C.-K.}\ \bibnamefont {Chiu}},
  \bibinfo {author} {\bibfnamefont {M.~Y.}\ \bibnamefont {Chou}}, \ and\
  \bibinfo {author} {\bibfnamefont {A.~P.}\ \bibnamefont {Schnyder}},\
  }\href@noop {} {\bibfield  {journal} {\bibinfo  {journal} {Phys. Rev. B}\
  }\textbf {\bibinfo {volume} {93}},\ \bibinfo {pages} {205132} (\bibinfo
  {year} {2016})}\BibitemShut {NoStop}%
\bibitem [{\citenamefont {Schoop}\ \emph {et~al.}(2016)\citenamefont {Schoop},
  \citenamefont {Ali}, \citenamefont {Straßer}, \citenamefont {Topp},
  \citenamefont {Varykhalov}, \citenamefont {Marchenko}, \citenamefont
  {Duppel}, \citenamefont {Parkin}, \citenamefont {Lotsch},\ and\ \citenamefont
  {Ast}}]{33}%
  \BibitemOpen
  \bibfield  {author} {\bibinfo {author} {\bibfnamefont {L.~M.}\ \bibnamefont
  {Schoop}}, \bibinfo {author} {\bibfnamefont {M.~N.}\ \bibnamefont {Ali}},
  \bibinfo {author} {\bibfnamefont {C.}~\bibnamefont {Straßer}}, \bibinfo
  {author} {\bibfnamefont {A.}~\bibnamefont {Topp}}, \bibinfo {author}
  {\bibfnamefont {A.}~\bibnamefont {Varykhalov}}, \bibinfo {author}
  {\bibfnamefont {D.}~\bibnamefont {Marchenko}}, \bibinfo {author}
  {\bibfnamefont {V.}~\bibnamefont {Duppel}}, \bibinfo {author} {\bibfnamefont
  {S.~S.~P.}\ \bibnamefont {Parkin}}, \bibinfo {author} {\bibfnamefont {B.~V.}\
  \bibnamefont {Lotsch}}, \ and\ \bibinfo {author} {\bibfnamefont {C.~R.}\
  \bibnamefont {Ast}},\ }\href@noop {} {\bibfield  {journal} {\bibinfo
  {journal} {Nat. Commun.}\ }\textbf {\bibinfo {volume} {7}},\ \bibinfo {pages}
  {11696} (\bibinfo {year} {2016})}\BibitemShut {NoStop}%
\bibitem [{\citenamefont {Liu}\ \emph {et~al.}(2016{\natexlab{b}})\citenamefont
  {Liu}, \citenamefont {Kriegner}, \citenamefont {Horak}, \citenamefont
  {Puggioni}, \citenamefont {Rayan~Serrao}, \citenamefont {Chen}, \citenamefont
  {Yi}, \citenamefont {Frontera}, \citenamefont {Holy}, \citenamefont
  {Vishwanath}, \citenamefont {Rondinelli}, \citenamefont {Marti},\ and\
  \citenamefont {Ramesh}}]{34}%
  \BibitemOpen
  \bibfield  {author} {\bibinfo {author} {\bibfnamefont {J.}~\bibnamefont
  {Liu}}, \bibinfo {author} {\bibfnamefont {D.}~\bibnamefont {Kriegner}},
  \bibinfo {author} {\bibfnamefont {L.}~\bibnamefont {Horak}}, \bibinfo
  {author} {\bibfnamefont {D.}~\bibnamefont {Puggioni}}, \bibinfo {author}
  {\bibfnamefont {C.}~\bibnamefont {Rayan~Serrao}}, \bibinfo {author}
  {\bibfnamefont {R.}~\bibnamefont {Chen}}, \bibinfo {author} {\bibfnamefont
  {D.}~\bibnamefont {Yi}}, \bibinfo {author} {\bibfnamefont {C.}~\bibnamefont
  {Frontera}}, \bibinfo {author} {\bibfnamefont {V.}~\bibnamefont {Holy}},
  \bibinfo {author} {\bibfnamefont {A.}~\bibnamefont {Vishwanath}}, \bibinfo
  {author} {\bibfnamefont {J.~M.}\ \bibnamefont {Rondinelli}}, \bibinfo
  {author} {\bibfnamefont {X.}~\bibnamefont {Marti}}, \ and\ \bibinfo {author}
  {\bibfnamefont {R.}~\bibnamefont {Ramesh}},\ }\href@noop {} {\bibfield
  {journal} {\bibinfo  {journal} {Phys. Rev. B}\ }\textbf {\bibinfo {volume}
  {93}},\ \bibinfo {pages} {085118} (\bibinfo {year}
  {2016}{\natexlab{b}})}\BibitemShut {NoStop}%
\bibitem [{\citenamefont {Liang}\ \emph {et~al.}(2016)\citenamefont {Liang},
  \citenamefont {Zhou}, \citenamefont {Yu}, \citenamefont {Wang},\ and\
  \citenamefont {Weng}}]{35}%
  \BibitemOpen
  \bibfield  {author} {\bibinfo {author} {\bibfnamefont {Q.-F.}\ \bibnamefont
  {Liang}}, \bibinfo {author} {\bibfnamefont {J.}~\bibnamefont {Zhou}},
  \bibinfo {author} {\bibfnamefont {R.}~\bibnamefont {Yu}}, \bibinfo {author}
  {\bibfnamefont {Z.}~\bibnamefont {Wang}}, \ and\ \bibinfo {author}
  {\bibfnamefont {H.}~\bibnamefont {Weng}},\ }\href@noop {} {\bibfield
  {journal} {\bibinfo  {journal} {Phys. Rev. B}\ }\textbf {\bibinfo {volume}
  {93}},\ \bibinfo {pages} {085427} (\bibinfo {year} {2016})}\BibitemShut
  {NoStop}%
\bibitem [{\citenamefont {Li}\ \emph {et~al.}(2016)\citenamefont {Li},
  \citenamefont {Ma}, \citenamefont {Cheng}, \citenamefont {Wang},
  \citenamefont {Li}, \citenamefont {Zhang}, \citenamefont {Li},\ and\
  \citenamefont {Chen}}]{36}%
  \BibitemOpen
  \bibfield  {author} {\bibinfo {author} {\bibfnamefont {R.}~\bibnamefont
  {Li}}, \bibinfo {author} {\bibfnamefont {H.}~\bibnamefont {Ma}}, \bibinfo
  {author} {\bibfnamefont {X.}~\bibnamefont {Cheng}}, \bibinfo {author}
  {\bibfnamefont {S.}~\bibnamefont {Wang}}, \bibinfo {author} {\bibfnamefont
  {D.}~\bibnamefont {Li}}, \bibinfo {author} {\bibfnamefont {Z.}~\bibnamefont
  {Zhang}}, \bibinfo {author} {\bibfnamefont {Y.}~\bibnamefont {Li}}, \ and\
  \bibinfo {author} {\bibfnamefont {X.-Q.}\ \bibnamefont {Chen}},\ }\href@noop
  {} {\bibfield  {journal} {\bibinfo  {journal} {Phys. Rev. Lett.}\ }\textbf
  {\bibinfo {volume} {117}},\ \bibinfo {pages} {096401} (\bibinfo {year}
  {2016})}\BibitemShut {NoStop}%
\bibitem [{\citenamefont {Kobayashi}\ \emph {et~al.}(2017)\citenamefont
  {Kobayashi}, \citenamefont {Yamakawa}, \citenamefont {Yamakage},
  \citenamefont {Inohara}, \citenamefont {Okamoto},\ and\ \citenamefont
  {Tanaka}}]{37}%
  \BibitemOpen
  \bibfield  {author} {\bibinfo {author} {\bibfnamefont {S.}~\bibnamefont
  {Kobayashi}}, \bibinfo {author} {\bibfnamefont {Y.}~\bibnamefont {Yamakawa}},
  \bibinfo {author} {\bibfnamefont {A.}~\bibnamefont {Yamakage}}, \bibinfo
  {author} {\bibfnamefont {T.}~\bibnamefont {Inohara}}, \bibinfo {author}
  {\bibfnamefont {Y.}~\bibnamefont {Okamoto}}, \ and\ \bibinfo {author}
  {\bibfnamefont {Y.}~\bibnamefont {Tanaka}},\ }\href@noop {} {\bibfield
  {journal} {\bibinfo  {journal} {Phys. Rev. B}\ }\textbf {\bibinfo {volume}
  {95}},\ \bibinfo {pages} {245208} (\bibinfo {year} {2017})}\BibitemShut
  {NoStop}%
\bibitem [{\citenamefont {Sun}\ \emph {et~al.}(2017)\citenamefont {Sun},
  \citenamefont {Zhang}, \citenamefont {Liu}, \citenamefont {Felser},\ and\
  \citenamefont {Yan}}]{38}%
  \BibitemOpen
  \bibfield  {author} {\bibinfo {author} {\bibfnamefont {Y.}~\bibnamefont
  {Sun}}, \bibinfo {author} {\bibfnamefont {Y.}~\bibnamefont {Zhang}}, \bibinfo
  {author} {\bibfnamefont {C.-X.}\ \bibnamefont {Liu}}, \bibinfo {author}
  {\bibfnamefont {C.}~\bibnamefont {Felser}}, \ and\ \bibinfo {author}
  {\bibfnamefont {B.}~\bibnamefont {Yan}},\ }\href@noop {} {\bibfield
  {journal} {\bibinfo  {journal} {Phys. Rev. B}\ }\textbf {\bibinfo {volume}
  {95}},\ \bibinfo {pages} {235104} (\bibinfo {year} {2017})}\BibitemShut
  {NoStop}%
\bibitem [{\citenamefont {Li}\ \emph {et~al.}(2017{\natexlab{a}})\citenamefont
  {Li}, \citenamefont {Yu}, \citenamefont {Liu}, \citenamefont {Guan},
  \citenamefont {Wang}, \citenamefont {Zhang}, \citenamefont {Yao},\ and\
  \citenamefont {Yang}}]{39}%
  \BibitemOpen
  \bibfield  {author} {\bibinfo {author} {\bibfnamefont {S.}~\bibnamefont
  {Li}}, \bibinfo {author} {\bibfnamefont {Z.-M.}\ \bibnamefont {Yu}}, \bibinfo
  {author} {\bibfnamefont {Y.}~\bibnamefont {Liu}}, \bibinfo {author}
  {\bibfnamefont {S.}~\bibnamefont {Guan}}, \bibinfo {author} {\bibfnamefont
  {S.-S.}\ \bibnamefont {Wang}}, \bibinfo {author} {\bibfnamefont
  {X.}~\bibnamefont {Zhang}}, \bibinfo {author} {\bibfnamefont
  {Y.}~\bibnamefont {Yao}}, \ and\ \bibinfo {author} {\bibfnamefont {S.~A.}\
  \bibnamefont {Yang}},\ }\href@noop {} {\bibfield  {journal} {\bibinfo
  {journal} {Phys. Rev. B}\ }\textbf {\bibinfo {volume} {96}},\ \bibinfo
  {pages} {081106} (\bibinfo {year} {2017}{\natexlab{a}})}\BibitemShut
  {NoStop}%
\bibitem [{\citenamefont {Wang}(2017)}]{40}%
  \BibitemOpen
  \bibfield  {author} {\bibinfo {author} {\bibfnamefont {J.}~\bibnamefont
  {Wang}},\ }\href@noop {} {\bibfield  {journal} {\bibinfo  {journal} {Phys.
  Rev. B}\ }\textbf {\bibinfo {volume} {96}},\ \bibinfo {pages} {081107}
  (\bibinfo {year} {2017})}\BibitemShut {NoStop}%
\bibitem [{\citenamefont {Gong}\ \emph {et~al.}(2018)\citenamefont {Gong},
  \citenamefont {Xie}, \citenamefont {Chen}, \citenamefont {Kim},\ and\
  \citenamefont {Vanderbilt}}]{41}%
  \BibitemOpen
  \bibfield  {author} {\bibinfo {author} {\bibfnamefont {C.}~\bibnamefont
  {Gong}}, \bibinfo {author} {\bibfnamefont {Y.}~\bibnamefont {Xie}}, \bibinfo
  {author} {\bibfnamefont {Y.}~\bibnamefont {Chen}}, \bibinfo {author}
  {\bibfnamefont {H.-S.}\ \bibnamefont {Kim}}, \ and\ \bibinfo {author}
  {\bibfnamefont {D.}~\bibnamefont {Vanderbilt}},\ }\href@noop {} {\bibfield
  {journal} {\bibinfo  {journal} {Physical Review Letters}\ }\textbf {\bibinfo
  {volume} {120}},\ \bibinfo {pages} {106403} (\bibinfo {year}
  {2018})}\BibitemShut {NoStop}%
\bibitem [{\citenamefont {Bzdusek}\ \emph {et~al.}(2016)\citenamefont
  {Bzdusek}, \citenamefont {Wu}, \citenamefont {Ruegg}, \citenamefont
  {Sigrist}, \citenamefont {Soluyanov} \emph {et~al.}}]{42}%
  \BibitemOpen
  \bibfield  {author} {\bibinfo {author} {\bibfnamefont {T.}~\bibnamefont
  {Bzdusek}}, \bibinfo {author} {\bibfnamefont {Q.}~\bibnamefont {Wu}},
  \bibinfo {author} {\bibfnamefont {A.}~\bibnamefont {Ruegg}}, \bibinfo
  {author} {\bibfnamefont {M.}~\bibnamefont {Sigrist}}, \bibinfo {author}
  {\bibfnamefont {A.}~\bibnamefont {Soluyanov}},  \emph {et~al.},\ }\href@noop
  {} {\bibfield  {journal} {\bibinfo  {journal} {Nature}\ }\textbf {\bibinfo
  {volume} {538}},\ \bibinfo {pages} {75} (\bibinfo {year} {2016})}\BibitemShut
  {NoStop}%
\bibitem [{\citenamefont {Yu}\ \emph {et~al.}(2017)\citenamefont {Yu},
  \citenamefont {Wu}, \citenamefont {Fang},\ and\ \citenamefont {Weng}}]{43}%
  \BibitemOpen
  \bibfield  {author} {\bibinfo {author} {\bibfnamefont {R.}~\bibnamefont
  {Yu}}, \bibinfo {author} {\bibfnamefont {Q.}~\bibnamefont {Wu}}, \bibinfo
  {author} {\bibfnamefont {Z.}~\bibnamefont {Fang}}, \ and\ \bibinfo {author}
  {\bibfnamefont {H.}~\bibnamefont {Weng}},\ }\href@noop {} {\bibfield
  {journal} {\bibinfo  {journal} {Phys. Rev. Lett.}\ }\textbf {\bibinfo
  {volume} {119}},\ \bibinfo {pages} {036401} (\bibinfo {year}
  {2017})}\BibitemShut {NoStop}%
\bibitem [{\citenamefont {Yan}\ \emph {et~al.}(2017)\citenamefont {Yan},
  \citenamefont {Bi}, \citenamefont {Shen}, \citenamefont {Lu}, \citenamefont
  {Zhang},\ and\ \citenamefont {Wang}}]{44}%
  \BibitemOpen
  \bibfield  {author} {\bibinfo {author} {\bibfnamefont {Z.}~\bibnamefont
  {Yan}}, \bibinfo {author} {\bibfnamefont {R.}~\bibnamefont {Bi}}, \bibinfo
  {author} {\bibfnamefont {H.}~\bibnamefont {Shen}}, \bibinfo {author}
  {\bibfnamefont {L.}~\bibnamefont {Lu}}, \bibinfo {author} {\bibfnamefont
  {S.-C.}\ \bibnamefont {Zhang}}, \ and\ \bibinfo {author} {\bibfnamefont
  {Z.}~\bibnamefont {Wang}},\ }\href@noop {} {\bibfield  {journal} {\bibinfo
  {journal} {Phys. Rev. B}\ }\textbf {\bibinfo {volume} {96}},\ \bibinfo
  {pages} {041103} (\bibinfo {year} {2017})}\BibitemShut {NoStop}%
\bibitem [{\citenamefont {Chen}\ \emph {et~al.}(2017)\citenamefont {Chen},
  \citenamefont {Lu},\ and\ \citenamefont {Hou}}]{45}%
  \BibitemOpen
  \bibfield  {author} {\bibinfo {author} {\bibfnamefont {W.}~\bibnamefont
  {Chen}}, \bibinfo {author} {\bibfnamefont {H.-Z.}\ \bibnamefont {Lu}}, \ and\
  \bibinfo {author} {\bibfnamefont {J.-M.}\ \bibnamefont {Hou}},\ }\href@noop
  {} {\bibfield  {journal} {\bibinfo  {journal} {Phys. Rev. B}\ }\textbf
  {\bibinfo {volume} {96}},\ \bibinfo {pages} {041102} (\bibinfo {year}
  {2017})}\BibitemShut {NoStop}%
\bibitem [{\citenamefont {Ezawa}(2017)}]{46}%
  \BibitemOpen
  \bibfield  {author} {\bibinfo {author} {\bibfnamefont {M.}~\bibnamefont
  {Ezawa}},\ }\href@noop {} {\bibfield  {journal} {\bibinfo  {journal} {Phys.
  Rev. B}\ }\textbf {\bibinfo {volume} {96}},\ \bibinfo {pages} {041202}
  (\bibinfo {year} {2017})}\BibitemShut {NoStop}%
\bibitem [{\citenamefont {Feng}\ \emph {et~al.}(2017)\citenamefont {Feng},
  \citenamefont {Yue}, \citenamefont {Song}, \citenamefont {Wu},\ and\
  \citenamefont {Wen}}]{47}%
  \BibitemOpen
  \bibfield  {author} {\bibinfo {author} {\bibfnamefont {X.}~\bibnamefont
  {Feng}}, \bibinfo {author} {\bibfnamefont {C.}~\bibnamefont {Yue}}, \bibinfo
  {author} {\bibfnamefont {Z.}~\bibnamefont {Song}}, \bibinfo {author}
  {\bibfnamefont {Q.}~\bibnamefont {Wu}}, \ and\ \bibinfo {author}
  {\bibfnamefont {B.}~\bibnamefont {Wen}},\ }\href@noop {} {\bibfield
  {journal} {\bibinfo  {journal} {arXiv preprint arXiv:1705.00511}\ } (\bibinfo
  {year} {2017})}\BibitemShut {NoStop}%
\bibitem [{\citenamefont {Jiao}\ \emph {et~al.}(2017)\citenamefont {Jiao},
  \citenamefont {Ma}, \citenamefont {Zhang}, \citenamefont {Bell},
  \citenamefont {Sanvito},\ and\ \citenamefont {Du}}]{48}%
  \BibitemOpen
  \bibfield  {author} {\bibinfo {author} {\bibfnamefont {Y.}~\bibnamefont
  {Jiao}}, \bibinfo {author} {\bibfnamefont {F.}~\bibnamefont {Ma}}, \bibinfo
  {author} {\bibfnamefont {C.}~\bibnamefont {Zhang}}, \bibinfo {author}
  {\bibfnamefont {J.}~\bibnamefont {Bell}}, \bibinfo {author} {\bibfnamefont
  {S.}~\bibnamefont {Sanvito}}, \ and\ \bibinfo {author} {\bibfnamefont
  {A.}~\bibnamefont {Du}},\ }\href@noop {} {\bibfield  {journal} {\bibinfo
  {journal} {Phys. Rev. Lett.}\ }\textbf {\bibinfo {volume} {119}},\ \bibinfo
  {pages} {016403} (\bibinfo {year} {2017})}\BibitemShut {NoStop}%
\bibitem [{\citenamefont {Bi}\ \emph {et~al.}(2017)\citenamefont {Bi},
  \citenamefont {Yan}, \citenamefont {Lu},\ and\ \citenamefont {Wang}}]{49}%
  \BibitemOpen
  \bibfield  {author} {\bibinfo {author} {\bibfnamefont {R.}~\bibnamefont
  {Bi}}, \bibinfo {author} {\bibfnamefont {Z.}~\bibnamefont {Yan}}, \bibinfo
  {author} {\bibfnamefont {L.}~\bibnamefont {Lu}}, \ and\ \bibinfo {author}
  {\bibfnamefont {Z.}~\bibnamefont {Wang}},\ }\href@noop {} {\bibfield
  {journal} {\bibinfo  {journal} {Phys. Rev. B}\ }\textbf {\bibinfo {volume}
  {96}},\ \bibinfo {pages} {201305} (\bibinfo {year} {2017})}\BibitemShut
  {NoStop}%
\bibitem [{\citenamefont {Wang}\ \emph {et~al.}(2018)\citenamefont {Wang},
  \citenamefont {Nie}, \citenamefont {Weng}, \citenamefont {Kawazoe},\ and\
  \citenamefont {Chen}}]{50}%
  \BibitemOpen
  \bibfield  {author} {\bibinfo {author} {\bibfnamefont {J.-T.}\ \bibnamefont
  {Wang}}, \bibinfo {author} {\bibfnamefont {S.}~\bibnamefont {Nie}}, \bibinfo
  {author} {\bibfnamefont {H.}~\bibnamefont {Weng}}, \bibinfo {author}
  {\bibfnamefont {Y.}~\bibnamefont {Kawazoe}}, \ and\ \bibinfo {author}
  {\bibfnamefont {C.}~\bibnamefont {Chen}},\ }\href@noop {} {\bibfield
  {journal} {\bibinfo  {journal} {Phys. Rev. Lett.}\ }\textbf {\bibinfo
  {volume} {120}},\ \bibinfo {pages} {026402} (\bibinfo {year}
  {2018})}\BibitemShut {NoStop}%
\bibitem [{\citenamefont {Liang}\ \emph {et~al.}(2015)\citenamefont {Liang},
  \citenamefont {Gibson}, \citenamefont {Ali}, \citenamefont {Liu},
  \citenamefont {Cava},\ and\ \citenamefont {Ong}}]{51}%
  \BibitemOpen
  \bibfield  {author} {\bibinfo {author} {\bibfnamefont {T.}~\bibnamefont
  {Liang}}, \bibinfo {author} {\bibfnamefont {Q.}~\bibnamefont {Gibson}},
  \bibinfo {author} {\bibfnamefont {M.~N.}\ \bibnamefont {Ali}}, \bibinfo
  {author} {\bibfnamefont {M.}~\bibnamefont {Liu}}, \bibinfo {author}
  {\bibfnamefont {R.}~\bibnamefont {Cava}}, \ and\ \bibinfo {author}
  {\bibfnamefont {N.}~\bibnamefont {Ong}},\ }\href@noop {} {\bibfield
  {journal} {\bibinfo  {journal} {Nat. Mater.}\ }\textbf {\bibinfo {volume}
  {14}},\ \bibinfo {pages} {280} (\bibinfo {year} {2015})}\BibitemShut
  {NoStop}%
\bibitem [{\citenamefont {Huh}\ \emph {et~al.}(2016)\citenamefont {Huh},
  \citenamefont {Moon},\ and\ \citenamefont {Kim}}]{52}%
  \BibitemOpen
  \bibfield  {author} {\bibinfo {author} {\bibfnamefont {Y.}~\bibnamefont
  {Huh}}, \bibinfo {author} {\bibfnamefont {E.-G.}\ \bibnamefont {Moon}}, \
  and\ \bibinfo {author} {\bibfnamefont {Y.~B.}\ \bibnamefont {Kim}},\
  }\href@noop {} {\bibfield  {journal} {\bibinfo  {journal} {Phys. Rev. B}\
  }\textbf {\bibinfo {volume} {93}},\ \bibinfo {pages} {035138} (\bibinfo
  {year} {2016})}\BibitemShut {NoStop}%
\bibitem [{\citenamefont {Rhim}\ and\ \citenamefont {Kim}(2015)}]{53}%
  \BibitemOpen
  \bibfield  {author} {\bibinfo {author} {\bibfnamefont {J.-W.}\ \bibnamefont
  {Rhim}}\ and\ \bibinfo {author} {\bibfnamefont {Y.~B.}\ \bibnamefont {Kim}},\
  }\href@noop {} {\bibfield  {journal} {\bibinfo  {journal} {Phys. Rev. B}\
  }\textbf {\bibinfo {volume} {92}},\ \bibinfo {pages} {045126} (\bibinfo
  {year} {2015})}\BibitemShut {NoStop}%
\bibitem [{\citenamefont {Yan}\ \emph {et~al.}(2016)\citenamefont {Yan},
  \citenamefont {Huang},\ and\ \citenamefont {Wang}}]{54}%
  \BibitemOpen
  \bibfield  {author} {\bibinfo {author} {\bibfnamefont {Z.}~\bibnamefont
  {Yan}}, \bibinfo {author} {\bibfnamefont {P.-W.}\ \bibnamefont {Huang}}, \
  and\ \bibinfo {author} {\bibfnamefont {Z.}~\bibnamefont {Wang}},\ }\href@noop
  {} {\bibfield  {journal} {\bibinfo  {journal} {Phys. Rev. B}\ }\textbf
  {\bibinfo {volume} {93}},\ \bibinfo {pages} {085138} (\bibinfo {year}
  {2016})}\BibitemShut {NoStop}%
\bibitem [{\citenamefont {Harada}\ \emph {et~al.}(1981)\citenamefont {Harada},
  \citenamefont {Sasa},\ and\ \citenamefont {Uda}}]{55}%
  \BibitemOpen
  \bibfield  {author} {\bibinfo {author} {\bibfnamefont {H.}~\bibnamefont
  {Harada}}, \bibinfo {author} {\bibfnamefont {Y.}~\bibnamefont {Sasa}}, \ and\
  \bibinfo {author} {\bibfnamefont {M.}~\bibnamefont {Uda}},\ }\href@noop {}
  {\bibfield  {journal} {\bibinfo  {journal} {J. Appl. Crystallogr.}\ }\textbf
  {\bibinfo {volume} {14}},\ \bibinfo {pages} {141} (\bibinfo {year}
  {1981})}\BibitemShut {NoStop}%
\bibitem [{\citenamefont {Syono}\ and\ \citenamefont {Akimoto}(1968)}]{56}%
  \BibitemOpen
  \bibfield  {author} {\bibinfo {author} {\bibfnamefont {Y.}~\bibnamefont
  {Syono}}\ and\ \bibinfo {author} {\bibfnamefont {S.}~\bibnamefont
  {Akimoto}},\ }\href@noop {} {\bibfield  {journal} {\bibinfo  {journal}
  {Mater. Res. Bull.}\ }\textbf {\bibinfo {volume} {3}},\ \bibinfo {pages}
  {153} (\bibinfo {year} {1968})}\BibitemShut {NoStop}%
\bibitem [{\citenamefont {Hill}(1982)}]{57}%
  \BibitemOpen
  \bibfield  {author} {\bibinfo {author} {\bibfnamefont {R.~J.}\ \bibnamefont
  {Hill}},\ }\href@noop {} {\bibfield  {journal} {\bibinfo  {journal} {Mater.
  Res. Bull.}\ }\textbf {\bibinfo {volume} {17}},\ \bibinfo {pages} {769}
  (\bibinfo {year} {1982})}\BibitemShut {NoStop}%
\bibitem [{\citenamefont {Heinemann}\ \emph {et~al.}(1995)\citenamefont
  {Heinemann}, \citenamefont {Terpstra}, \citenamefont {Haas},\ and\
  \citenamefont {De~Groot}}]{58}%
  \BibitemOpen
  \bibfield  {author} {\bibinfo {author} {\bibfnamefont {M.}~\bibnamefont
  {Heinemann}}, \bibinfo {author} {\bibfnamefont {H.}~\bibnamefont {Terpstra}},
  \bibinfo {author} {\bibfnamefont {C.}~\bibnamefont {Haas}}, \ and\ \bibinfo
  {author} {\bibfnamefont {R.}~\bibnamefont {De~Groot}},\ }\href@noop {}
  {\bibfield  {journal} {\bibinfo  {journal} {Phys. Rev. B}\ }\textbf {\bibinfo
  {volume} {52}},\ \bibinfo {pages} {11740} (\bibinfo {year}
  {1995})}\BibitemShut {NoStop}%
\bibitem [{\citenamefont {Payne}\ \emph {et~al.}(2005)\citenamefont {Payne},
  \citenamefont {Egdell}, \citenamefont {Hao}, \citenamefont {Foord},
  \citenamefont {Walsh},\ and\ \citenamefont {Watson}}]{59}%
  \BibitemOpen
  \bibfield  {author} {\bibinfo {author} {\bibfnamefont {D.~J.}\ \bibnamefont
  {Payne}}, \bibinfo {author} {\bibfnamefont {R.~G.}\ \bibnamefont {Egdell}},
  \bibinfo {author} {\bibfnamefont {W.}~\bibnamefont {Hao}}, \bibinfo {author}
  {\bibfnamefont {J.~S.}\ \bibnamefont {Foord}}, \bibinfo {author}
  {\bibfnamefont {A.}~\bibnamefont {Walsh}}, \ and\ \bibinfo {author}
  {\bibfnamefont {G.~W.}\ \bibnamefont {Watson}},\ }\href@noop {} {\bibfield
  {journal} {\bibinfo  {journal} {Chem. Phys. Lett.}\ }\textbf {\bibinfo
  {volume} {411}},\ \bibinfo {pages} {181} (\bibinfo {year}
  {2005})}\BibitemShut {NoStop}%
\bibitem [{\citenamefont {Payne}\ \emph {et~al.}(2007)\citenamefont {Payne},
  \citenamefont {Egdell}, \citenamefont {Paolicelli}, \citenamefont {Offi},
  \citenamefont {Panaccione}, \citenamefont {Lacovig}, \citenamefont {Monaco},
  \citenamefont {Vanko}, \citenamefont {Walsh}, \citenamefont {Watson} \emph
  {et~al.}}]{60}%
  \BibitemOpen
  \bibfield  {author} {\bibinfo {author} {\bibfnamefont {D.}~\bibnamefont
  {Payne}}, \bibinfo {author} {\bibfnamefont {R.}~\bibnamefont {Egdell}},
  \bibinfo {author} {\bibfnamefont {G.}~\bibnamefont {Paolicelli}}, \bibinfo
  {author} {\bibfnamefont {F.}~\bibnamefont {Offi}}, \bibinfo {author}
  {\bibfnamefont {G.}~\bibnamefont {Panaccione}}, \bibinfo {author}
  {\bibfnamefont {P.}~\bibnamefont {Lacovig}}, \bibinfo {author} {\bibfnamefont
  {G.}~\bibnamefont {Monaco}}, \bibinfo {author} {\bibfnamefont
  {G.}~\bibnamefont {Vanko}}, \bibinfo {author} {\bibfnamefont
  {A.}~\bibnamefont {Walsh}}, \bibinfo {author} {\bibfnamefont
  {G.}~\bibnamefont {Watson}},  \emph {et~al.},\ }\href@noop {} {\bibfield
  {journal} {\bibinfo  {journal} {Phys. Rev. B}\ }\textbf {\bibinfo {volume}
  {75}},\ \bibinfo {pages} {153102} (\bibinfo {year} {2007})}\BibitemShut
  {NoStop}%
\bibitem [{\citenamefont {Scanlon}\ \emph {et~al.}(2011)\citenamefont
  {Scanlon}, \citenamefont {Kehoe}, \citenamefont {Watson}, \citenamefont
  {Jones}, \citenamefont {David}, \citenamefont {Payne}, \citenamefont
  {Egdell}, \citenamefont {Edwards},\ and\ \citenamefont {Walsh}}]{61}%
  \BibitemOpen
  \bibfield  {author} {\bibinfo {author} {\bibfnamefont {D.~O.}\ \bibnamefont
  {Scanlon}}, \bibinfo {author} {\bibfnamefont {A.~B.}\ \bibnamefont {Kehoe}},
  \bibinfo {author} {\bibfnamefont {G.~W.}\ \bibnamefont {Watson}}, \bibinfo
  {author} {\bibfnamefont {M.~O.}\ \bibnamefont {Jones}}, \bibinfo {author}
  {\bibfnamefont {W.~I.}\ \bibnamefont {David}}, \bibinfo {author}
  {\bibfnamefont {D.~J.}\ \bibnamefont {Payne}}, \bibinfo {author}
  {\bibfnamefont {R.~G.}\ \bibnamefont {Egdell}}, \bibinfo {author}
  {\bibfnamefont {P.~P.}\ \bibnamefont {Edwards}}, \ and\ \bibinfo {author}
  {\bibfnamefont {A.}~\bibnamefont {Walsh}},\ }\href@noop {} {\bibfield
  {journal} {\bibinfo  {journal} {Phys. Rev. Lett.}\ }\textbf {\bibinfo
  {volume} {107}},\ \bibinfo {pages} {246402} (\bibinfo {year}
  {2011})}\BibitemShut {NoStop}%
\bibitem [{\citenamefont {Walsh}\ \emph {et~al.}(2013)\citenamefont {Walsh},
  \citenamefont {Kehoe}, \citenamefont {Temple}, \citenamefont {Watson},\ and\
  \citenamefont {Scanlon}}]{62}%
  \BibitemOpen
  \bibfield  {author} {\bibinfo {author} {\bibfnamefont {A.}~\bibnamefont
  {Walsh}}, \bibinfo {author} {\bibfnamefont {A.~B.}\ \bibnamefont {Kehoe}},
  \bibinfo {author} {\bibfnamefont {D.~J.}\ \bibnamefont {Temple}}, \bibinfo
  {author} {\bibfnamefont {G.~W.}\ \bibnamefont {Watson}}, \ and\ \bibinfo
  {author} {\bibfnamefont {D.~O.}\ \bibnamefont {Scanlon}},\ }\href@noop {}
  {\bibfield  {journal} {\bibinfo  {journal} {Chem. Commun.}\ }\textbf
  {\bibinfo {volume} {49}},\ \bibinfo {pages} {448} (\bibinfo {year}
  {2013})}\BibitemShut {NoStop}%
\bibitem [{\citenamefont {Ma}\ \emph {et~al.}(2016)\citenamefont {Ma},
  \citenamefont {Jiao}, \citenamefont {Gao}, \citenamefont {Gu}, \citenamefont
  {Bilic}, \citenamefont {Sanvito},\ and\ \citenamefont {Du}}]{63}%
  \BibitemOpen
  \bibfield  {author} {\bibinfo {author} {\bibfnamefont {F.}~\bibnamefont
  {Ma}}, \bibinfo {author} {\bibfnamefont {Y.}~\bibnamefont {Jiao}}, \bibinfo
  {author} {\bibfnamefont {G.}~\bibnamefont {Gao}}, \bibinfo {author}
  {\bibfnamefont {Y.}~\bibnamefont {Gu}}, \bibinfo {author} {\bibfnamefont
  {A.}~\bibnamefont {Bilic}}, \bibinfo {author} {\bibfnamefont
  {S.}~\bibnamefont {Sanvito}}, \ and\ \bibinfo {author} {\bibfnamefont
  {A.}~\bibnamefont {Du}},\ }\href@noop {} {\bibfield  {journal} {\bibinfo
  {journal} {ACS Appl. Mater. Inter.}\ }\textbf {\bibinfo {volume} {8}},\
  \bibinfo {pages} {25667} (\bibinfo {year} {2016})}\BibitemShut {NoStop}%
\bibitem [{\citenamefont {Wang}\ and\ \citenamefont {Wang}(2017)}]{64}%
  \BibitemOpen
  \bibfield  {author} {\bibinfo {author} {\bibfnamefont {Z.}~\bibnamefont
  {Wang}}\ and\ \bibinfo {author} {\bibfnamefont {G.}~\bibnamefont {Wang}},\
  }\href@noop {} {\bibfield  {journal} {\bibinfo  {journal} {Phys. Lett. A}\
  }\textbf {\bibinfo {volume} {381}},\ \bibinfo {pages} {2856} (\bibinfo {year}
  {2017})}\BibitemShut {NoStop}%
\bibitem [{\citenamefont {Yu}\ \emph {et~al.}(2015)\citenamefont {Yu},
  \citenamefont {Weng}, \citenamefont {Fang}, \citenamefont {Dai},\ and\
  \citenamefont {Hu}}]{65}%
  \BibitemOpen
  \bibfield  {author} {\bibinfo {author} {\bibfnamefont {R.}~\bibnamefont
  {Yu}}, \bibinfo {author} {\bibfnamefont {H.}~\bibnamefont {Weng}}, \bibinfo
  {author} {\bibfnamefont {Z.}~\bibnamefont {Fang}}, \bibinfo {author}
  {\bibfnamefont {X.}~\bibnamefont {Dai}}, \ and\ \bibinfo {author}
  {\bibfnamefont {X.}~\bibnamefont {Hu}},\ }\href@noop {} {\bibfield  {journal}
  {\bibinfo  {journal} {Phys. Rev. Lett.}\ }\textbf {\bibinfo {volume} {115}},\
  \bibinfo {pages} {036807} (\bibinfo {year} {2015})}\BibitemShut {NoStop}%
\bibitem [{\citenamefont {Zeng}\ \emph {et~al.}(2015)\citenamefont {Zeng},
  \citenamefont {Fang}, \citenamefont {Chang}, \citenamefont {Chen},
  \citenamefont {Hsieh}, \citenamefont {Bansil}, \citenamefont {Lin},\ and\
  \citenamefont {Fu}}]{66}%
  \BibitemOpen
  \bibfield  {author} {\bibinfo {author} {\bibfnamefont {M.}~\bibnamefont
  {Zeng}}, \bibinfo {author} {\bibfnamefont {C.}~\bibnamefont {Fang}}, \bibinfo
  {author} {\bibfnamefont {G.}~\bibnamefont {Chang}}, \bibinfo {author}
  {\bibfnamefont {Y.-A.}\ \bibnamefont {Chen}}, \bibinfo {author}
  {\bibfnamefont {T.}~\bibnamefont {Hsieh}}, \bibinfo {author} {\bibfnamefont
  {A.}~\bibnamefont {Bansil}}, \bibinfo {author} {\bibfnamefont
  {H.}~\bibnamefont {Lin}}, \ and\ \bibinfo {author} {\bibfnamefont
  {L.}~\bibnamefont {Fu}},\ }\href@noop {} {\bibfield  {journal} {\bibinfo
  {journal} {arXiv preprint arXiv:1504.03492}\ } (\bibinfo {year}
  {2015})}\BibitemShut {NoStop}%
\bibitem [{\citenamefont {Du}\ \emph {et~al.}(2017)\citenamefont {Du},
  \citenamefont {Tang}, \citenamefont {Wang}, \citenamefont {Sheng},
  \citenamefont {Kan}, \citenamefont {Duan}, \citenamefont {Savrasov},\ and\
  \citenamefont {Wan}}]{67}%
  \BibitemOpen
  \bibfield  {author} {\bibinfo {author} {\bibfnamefont {Y.}~\bibnamefont
  {Du}}, \bibinfo {author} {\bibfnamefont {F.}~\bibnamefont {Tang}}, \bibinfo
  {author} {\bibfnamefont {D.}~\bibnamefont {Wang}}, \bibinfo {author}
  {\bibfnamefont {L.}~\bibnamefont {Sheng}}, \bibinfo {author} {\bibfnamefont
  {E.-j.}\ \bibnamefont {Kan}}, \bibinfo {author} {\bibfnamefont {C.-G.}\
  \bibnamefont {Duan}}, \bibinfo {author} {\bibfnamefont {S.~Y.}\ \bibnamefont
  {Savrasov}}, \ and\ \bibinfo {author} {\bibfnamefont {X.}~\bibnamefont
  {Wan}},\ }\href@noop {} {\bibfield  {journal} {\bibinfo  {journal} {npj
  Quantum Mater.}\ }\textbf {\bibinfo {volume} {2}},\ \bibinfo {pages} {3}
  (\bibinfo {year} {2017})}\BibitemShut {NoStop}%
\bibitem [{\citenamefont {Peng}\ \emph {et~al.}(2016)\citenamefont {Peng},
  \citenamefont {Yang}, \citenamefont {Perdew},\ and\ \citenamefont
  {Sun}}]{68}%
  \BibitemOpen
  \bibfield  {author} {\bibinfo {author} {\bibfnamefont {H.}~\bibnamefont
  {Peng}}, \bibinfo {author} {\bibfnamefont {Z.-H.}\ \bibnamefont {Yang}},
  \bibinfo {author} {\bibfnamefont {J.~P.}\ \bibnamefont {Perdew}}, \ and\
  \bibinfo {author} {\bibfnamefont {J.}~\bibnamefont {Sun}},\ }\href@noop {}
  {\bibfield  {journal} {\bibinfo  {journal} {Phys. Rev. X}\ }\textbf {\bibinfo
  {volume} {6}},\ \bibinfo {pages} {041005} (\bibinfo {year}
  {2016})}\BibitemShut {NoStop}%
\bibitem [{\citenamefont {Perdew}\ \emph {et~al.}(1996)\citenamefont {Perdew},
  \citenamefont {Burke},\ and\ \citenamefont {Ernzerhof}}]{69}%
  \BibitemOpen
  \bibfield  {author} {\bibinfo {author} {\bibfnamefont {J.~P.}\ \bibnamefont
  {Perdew}}, \bibinfo {author} {\bibfnamefont {K.}~\bibnamefont {Burke}}, \
  and\ \bibinfo {author} {\bibfnamefont {M.}~\bibnamefont {Ernzerhof}},\
  }\href@noop {} {\bibfield  {journal} {\bibinfo  {journal} {Phys. Rev. Lett.}\
  }\textbf {\bibinfo {volume} {77}},\ \bibinfo {pages} {3865} (\bibinfo {year}
  {1996})}\BibitemShut {NoStop}%
\bibitem [{\citenamefont {Bl\"ochl}(1994)}]{70}%
  \BibitemOpen
  \bibfield  {author} {\bibinfo {author} {\bibfnamefont {P.~E.}\ \bibnamefont
  {Bl\"ochl}},\ }\href@noop {} {\bibfield  {journal} {\bibinfo  {journal}
  {Phys. Rev. B}\ }\textbf {\bibinfo {volume} {50}},\ \bibinfo {pages} {17953}
  (\bibinfo {year} {1994})}\BibitemShut {NoStop}%
\bibitem [{\citenamefont {Heyd}\ \emph {et~al.}(2003)\citenamefont {Heyd},
  \citenamefont {Scuseria},\ and\ \citenamefont {Ernzerhof}}]{71}%
  \BibitemOpen
  \bibfield  {author} {\bibinfo {author} {\bibfnamefont {J.}~\bibnamefont
  {Heyd}}, \bibinfo {author} {\bibfnamefont {G.~E.}\ \bibnamefont {Scuseria}},
  \ and\ \bibinfo {author} {\bibfnamefont {M.}~\bibnamefont {Ernzerhof}},\
  }\href@noop {} {\bibfield  {journal} {\bibinfo  {journal} {J Chem. Phys.}\
  }\textbf {\bibinfo {volume} {118}},\ \bibinfo {pages} {8207} (\bibinfo {year}
  {2003})}\BibitemShut {NoStop}%
\bibitem [{\citenamefont {Heyd}\ \emph {et~al.}(2006)\citenamefont {Heyd},
  \citenamefont {Scuseria},\ and\ \citenamefont {Ernzerhof}}]{72}%
  \BibitemOpen
  \bibfield  {author} {\bibinfo {author} {\bibfnamefont {J.}~\bibnamefont
  {Heyd}}, \bibinfo {author} {\bibfnamefont {G.~E.}\ \bibnamefont {Scuseria}},
  \ and\ \bibinfo {author} {\bibfnamefont {M.}~\bibnamefont {Ernzerhof}},\
  }\href@noop {} {\bibfield  {journal} {\bibinfo  {journal} {J Chem. Phys.}\
  }\textbf {\bibinfo {volume} {124}},\ \bibinfo {pages} {219906} (\bibinfo
  {year} {2006})}\BibitemShut {NoStop}%
\bibitem [{\citenamefont {Mostofi}\ \emph {et~al.}(2008)\citenamefont
  {Mostofi}, \citenamefont {Yates}, \citenamefont {Lee}, \citenamefont {Souza},
  \citenamefont {Vanderbilt},\ and\ \citenamefont {Marzari}}]{73}%
  \BibitemOpen
  \bibfield  {author} {\bibinfo {author} {\bibfnamefont {A.~A.}\ \bibnamefont
  {Mostofi}}, \bibinfo {author} {\bibfnamefont {J.~R.}\ \bibnamefont {Yates}},
  \bibinfo {author} {\bibfnamefont {Y.-S.}\ \bibnamefont {Lee}}, \bibinfo
  {author} {\bibfnamefont {I.}~\bibnamefont {Souza}}, \bibinfo {author}
  {\bibfnamefont {D.}~\bibnamefont {Vanderbilt}}, \ and\ \bibinfo {author}
  {\bibfnamefont {N.}~\bibnamefont {Marzari}},\ }\href@noop {} {\bibfield
  {journal} {\bibinfo  {journal} {Comput. Phys. Commun.}\ }\textbf {\bibinfo
  {volume} {178}},\ \bibinfo {pages} {685 } (\bibinfo {year}
  {2008})}\BibitemShut {NoStop}%
\bibitem [{\citenamefont {Wu}\ \emph {et~al.}(2018)\citenamefont {Wu},
  \citenamefont {Zhang}, \citenamefont {Song}, \citenamefont {Troyer},\ and\
  \citenamefont {Soluyanov}}]{74}%
  \BibitemOpen
  \bibfield  {author} {\bibinfo {author} {\bibfnamefont {Q.}~\bibnamefont
  {Wu}}, \bibinfo {author} {\bibfnamefont {S.}~\bibnamefont {Zhang}}, \bibinfo
  {author} {\bibfnamefont {H.-F.}\ \bibnamefont {Song}}, \bibinfo {author}
  {\bibfnamefont {M.}~\bibnamefont {Troyer}}, \ and\ \bibinfo {author}
  {\bibfnamefont {A.~A.}\ \bibnamefont {Soluyanov}},\ }\href@noop {} {\bibfield
   {journal} {\bibinfo  {journal} {Comput. Phys. Commun.}\ }\textbf {\bibinfo
  {volume} {224}},\ \bibinfo {pages} {405} (\bibinfo {year}
  {2018})}\BibitemShut {NoStop}%
\bibitem [{\citenamefont {Huang}\ \emph
  {et~al.}(2016{\natexlab{b}})\citenamefont {Huang}, \citenamefont {Liu},
  \citenamefont {Vanderbilt},\ and\ \citenamefont {Duan}}]{75}%
  \BibitemOpen
  \bibfield  {author} {\bibinfo {author} {\bibfnamefont {H.}~\bibnamefont
  {Huang}}, \bibinfo {author} {\bibfnamefont {J.}~\bibnamefont {Liu}}, \bibinfo
  {author} {\bibfnamefont {D.}~\bibnamefont {Vanderbilt}}, \ and\ \bibinfo
  {author} {\bibfnamefont {W.}~\bibnamefont {Duan}},\ }\href@noop {} {\bibfield
   {journal} {\bibinfo  {journal} {Phys. Rev. B}\ }\textbf {\bibinfo {volume}
  {93}},\ \bibinfo {pages} {201114} (\bibinfo {year}
  {2016}{\natexlab{b}})}\BibitemShut {NoStop}%
\bibitem [{\citenamefont {Quan}\ \emph {et~al.}(2017)\citenamefont {Quan},
  \citenamefont {Yin},\ and\ \citenamefont {Pickett}}]{76}%
  \BibitemOpen
  \bibfield  {author} {\bibinfo {author} {\bibfnamefont {Y.}~\bibnamefont
  {Quan}}, \bibinfo {author} {\bibfnamefont {Z.}~\bibnamefont {Yin}}, \ and\
  \bibinfo {author} {\bibfnamefont {W.}~\bibnamefont {Pickett}},\ }\href@noop
  {} {\bibfield  {journal} {\bibinfo  {journal} {Phys. Rev. Lett.}\ }\textbf
  {\bibinfo {volume} {118}},\ \bibinfo {pages} {176402} (\bibinfo {year}
  {2017})}\BibitemShut {NoStop}%
\bibitem [{\citenamefont {Kim}\ \emph {et~al.}(2015{\natexlab{b}})\citenamefont
  {Kim}, \citenamefont {Wieder}, \citenamefont {Kane},\ and\ \citenamefont
  {Rappe}}]{77}%
  \BibitemOpen
  \bibfield  {author} {\bibinfo {author} {\bibfnamefont {Y.}~\bibnamefont
  {Kim}}, \bibinfo {author} {\bibfnamefont {B.~J.}\ \bibnamefont {Wieder}},
  \bibinfo {author} {\bibfnamefont {C.}~\bibnamefont {Kane}}, \ and\ \bibinfo
  {author} {\bibfnamefont {A.~M.}\ \bibnamefont {Rappe}},\ }\href@noop {}
  {\bibfield  {journal} {\bibinfo  {journal} {Phys. Rev. Lett.}\ }\textbf
  {\bibinfo {volume} {115}},\ \bibinfo {pages} {036806} (\bibinfo {year}
  {2015}{\natexlab{b}})}\BibitemShut {NoStop}%
\bibitem [{\citenamefont {Rajamathi}\ \emph {et~al.}(2017)\citenamefont
  {Rajamathi}, \citenamefont {Gupta}, \citenamefont {Kumar}, \citenamefont
  {Yang}, \citenamefont {Sun}, \citenamefont {S{\"u}{\ss}}, \citenamefont
  {Shekhar}, \citenamefont {Schmidt}, \citenamefont {Blumtritt}, \citenamefont
  {Werner} \emph {et~al.}}]{78}%
  \BibitemOpen
  \bibfield  {author} {\bibinfo {author} {\bibfnamefont {C.~R.}\ \bibnamefont
  {Rajamathi}}, \bibinfo {author} {\bibfnamefont {U.}~\bibnamefont {Gupta}},
  \bibinfo {author} {\bibfnamefont {N.}~\bibnamefont {Kumar}}, \bibinfo
  {author} {\bibfnamefont {H.}~\bibnamefont {Yang}}, \bibinfo {author}
  {\bibfnamefont {Y.}~\bibnamefont {Sun}}, \bibinfo {author} {\bibfnamefont
  {V.}~\bibnamefont {S{\"u}{\ss}}}, \bibinfo {author} {\bibfnamefont
  {C.}~\bibnamefont {Shekhar}}, \bibinfo {author} {\bibfnamefont
  {M.}~\bibnamefont {Schmidt}}, \bibinfo {author} {\bibfnamefont
  {H.}~\bibnamefont {Blumtritt}}, \bibinfo {author} {\bibfnamefont
  {P.}~\bibnamefont {Werner}},  \emph {et~al.},\ }\href@noop {} {\bibfield
  {journal} {\bibinfo  {journal} {Adv. Mater.}\ }\textbf {\bibinfo {volume}
  {29}} (\bibinfo {year} {2017})}\BibitemShut {NoStop}%
\bibitem [{\citenamefont {Li}\ \emph {et~al.}(2017{\natexlab{b}})\citenamefont
  {Li}, \citenamefont {Ma}, \citenamefont {Feng}, \citenamefont {Ullah},
  \citenamefont {Li}, \citenamefont {Dong}, \citenamefont {Li}, \citenamefont
  {Li},\ and\ \citenamefont {Chen}}]{79}%
  \BibitemOpen
  \bibfield  {author} {\bibinfo {author} {\bibfnamefont {J.}~\bibnamefont
  {Li}}, \bibinfo {author} {\bibfnamefont {H.}~\bibnamefont {Ma}}, \bibinfo
  {author} {\bibfnamefont {S.}~\bibnamefont {Feng}}, \bibinfo {author}
  {\bibfnamefont {S.}~\bibnamefont {Ullah}}, \bibinfo {author} {\bibfnamefont
  {R.}~\bibnamefont {Li}}, \bibinfo {author} {\bibfnamefont {J.}~\bibnamefont
  {Dong}}, \bibinfo {author} {\bibfnamefont {D.}~\bibnamefont {Li}}, \bibinfo
  {author} {\bibfnamefont {Y.}~\bibnamefont {Li}}, \ and\ \bibinfo {author}
  {\bibfnamefont {X.-Q.}\ \bibnamefont {Chen}},\ }\href@noop {} {\bibfield
  {journal} {\bibinfo  {journal} {arXiv preprint arXiv:1704.07043}\ } (\bibinfo
  {year} {2017}{\natexlab{b}})}\BibitemShut {NoStop}%
\end{thebibliography}%

\end{document}